\documentclass[12pt,letterpaper]{article}

\usepackage{verbatim}
\usepackage{amsmath}
\usepackage{amssymb}
\usepackage{graphicx}
\usepackage{cite}
\usepackage{subfig}
\usepackage{setspace}
\usepackage{url}
\usepackage[percent]{overpic}

\def\Re{{\cal R \mskip-4mu \lower.1ex \hbox{\it e}\,}}
\def\Im{{\cal I \mskip-5mu \lower.1ex \hbox{\it m}\,}}
\def\ie{{\it i.e.}}

\def\etc{{\it etc}}

\def\sub#1{_{\lower.25ex\hbox{$\scriptstyle#1$}}}

\def\to{\rightarrow}

\def\subw{_{\rm w}}
\def\mh{\ifmmode m\sbl H \else $m\sbl H$\fi}
\def\mch{\ifmmode m_{H^\pm} \else $m_{H^\pm}$\fi}
\def\mt{\ifmmode m_t\else $m_t$\fi}
\def\mc{\ifmmode m_c\else $m_c$\fi}
\def\mz{\ifmmode M_Z\else $M_Z$\fi}
\def\mw{\ifmmode M_W\else $M_W$\fi}
\def\mws{\ifmmode M_W^2 \else $M_W^2$\fi}
\def\mhs{\ifmmode m_H^2 \else $m_H^2$\fi}   
\def\mzs{\ifmmode M_Z^2 \else $M_Z^2$\fi}
\def\mts{\ifmmode m_t^2 \else $m_t^2$\fi}
\def\mcs{\ifmmode m_c^2 \else $m_c^2$\fi}
\def\mchs{\ifmmode m_{H^\pm}^2 \else $m_{H^\pm}^2$\fi}
\def\ztwo{\ifmmode Z_2\else $Z_2$\fi}
\def\zone{\ifmmode Z_1\else $Z_1$\fi}
\def\mtwo{\ifmmode M_2\else $M_2$\fi}
\def\mone{\ifmmode M_1\else $M_1$\fi}
\def\tb{\ifmmode \tan\beta \else $\tan\beta$\fi}
\def\xw{\ifmmode x\subw\else $x\subw$\fi}
\def\ch{\ifmmode H^\pm \else $H^\pm$\fi}
\def\lum{\ifmmode {\cal L}\else ${\cal L}$\fi}
\def\inpb{\ifmmode {\rm pb}^{-1}\else ${\rm pb}^{-1}$\fi}
\def\infb{\ifmmode {\rm fb}^{-1}\else ${\rm fb}^{-1}$\fi}
\def\epem{\ifmmode e^+e^-\else $e^+e^-$\fi}
\def\ppb{\ifmmode \bar pp\else $\bar pp$\fi}
\def\pbp{\ifmmode ~^(\bar p^)p\else $~^(\bar p^)p$\fi}
\def\bsg{\ifmmode B\to X_s\gamma\else $B\to latexilaX_s\gamma$\fi}
\def\bsll{\ifmmode B\to X_s\ell^+\ell^-\else $B\to X_s\ell^+\ell^-$\fi}
\def\bstt{\ifmmode B\to X_s\tau^+\tau^-\else $B\to X_s\tau^+\tau^-$\fi}

\newskip\zatskip \zatskip=0pt plus0pt minus0pt
\def\matth{\mathsurround=0pt}
\def\lsim{\mathrel{\mathpalette\atversim<}}

\def\atversim#1#2{\lower0.7ex\vbox{\baselineskip\zatskip\lineskip\zatskip
  \lineskiplimit 0pt\ialign{$\matth#1\hfil##\hfil$\crcr#2\crcr\sim\crcr}}}

\renewcommand{\thefootnote}{\fnsymbol{footnote}}

\begin{document} \begin{titlepage} 
\rightline{\vbox{\halign{&#\hfil\cr
&SLAC-PUB-15273\cr
}}}
\vspace{1in} 
\begin{center} 

{{\large\bf More Energy, More Searches, but the pMSSM Lives On}\footnote{Work supported by the Department of 
Energy, Contract DE-AC02-76SF00515}\\}

\medskip
\medskip
\normalsize 
{\large Matthew W.~Cahill-Rowley, JoAnne L.~Hewett, Ahmed Ismail, and

Thomas G.~Rizzo\footnote{email: mrowley, hewett, aismail, rizzo@slac.stanford.edu} \\
\vskip .6cm
SLAC National Accelerator Laboratory,  \\
2575 Sand Hill Rd, Menlo Park, CA 94025, USA\\}
\vskip .5cm

\end{center} 
\vskip 0.8cm

\begin{abstract} 

We further examine the capability of the 7 and 8 TeV LHC to explore the parameter space of the
p(henomenological)MSSM with neutralino LSPs.  Here we present an updated study employing all of the
relevant ATLAS SUSY analyses, as well as all relevant LHC non-MET searches, whose data were
publically available as of mid-September 2012.  We find that roughly 1/3 of our pMSSM model points, and 38\% of the points with colored sparticles below 1 TeV, are
excluded at present. An important role in excluding these models is being played by both the heavy flavor and multi-lepton
searches, as well as those for heavy stable charged particles.  Nonetheless, we find that light
gluinos, 1st/2nd generation squarks, and stop/sbottoms (of order 400-700 GeV), as well as
models with low fine-tuning (of order 1\% according to the Barbieri-Giudice measure), are still viable in the pMSSM.  In addition, we see that
increased luminosity at 8 TeV is unlikely to significantly improve the reach of the standard jets + MET searches. 
The impact of these null searches on the SUSY sparticle spectrum is discussed in detail and the 
implications of these results for models with low
fine-tuning, a future lepton collider and dark matter searches are examined.

\end{abstract}

\renewcommand{\thefootnote}{\arabic{footnote}} \end{titlepage}


\section{Introduction}
\label{sec:intro}

The LHC was designed to explore electroweak symmetry breaking and search for physics beyond the
Standard Model (SM).  The recent discovery of the Higgs boson~\cite{HiggsDiscovery} fulfills one of
these goals.  This discovery is a triumph for science as it profoundly deepens our understanding of the 
universe and completes our picture of the SM.  However, many fundamental questions remain outstanding, 
and it is imperative that the LHC experiments continue to search for new 
physics that may shed some insight or, possibly, open new doors of investigation.

To date, the onslaught of LHC data has challenged our most cherished theories of new physics~\cite{NP}.  
Search after search further constrains the parameter space of many models, with some scenarios
now being excluded.  In addition, the property of naturalness is now called into question.
Either new physics appears at the electroweak scale, or Nature has tuned the mass of the Higgs boson to one 
part in $\sim$10$^{32}$. 

Supersymmetry (SUSY) elegantly solves this hierarchy problem and is the most motivated and well-studied theory 
beyond the SM~\cite{SUSYrefs}. 
Naturalness considerations allow us to set limits on the values of parameters, given an upper bound on the level of fine-tuning permitted. In our recent work~\cite{CahillRowley:2012rv}, we showed that employing the Barbieri-Giudice definition~\cite{Ellis:1986yg, Barbieri:1987fn} of fine-tuning suggests the existence of at least one light stop (below $\sim$1 TeV), and implies that the Higgsino mass $|\mu|$ and stop mixing $|A_t|$ are below $\sim$456 GeV and $\sim$2.2 TeV, respectively, when requiring percent-level tuning or better. 
On the other hand, a Higgs mass of $\sim$126 GeV in the Minimal Supersymmetric Standard Model (MSSM) necessitates a large stop mass or large stop mixing to drive large radiative corrections.  This spells trouble for the simplest or constrained SUSY models, such as mSUGRA, which have difficulty satisfying all
experimental constraints (including the Higgs mass) while meeting the requirements for low fine-tuning. In particular, ATLAS and CMS have placed bounds~\cite{LHCcMSSM,ATLAS:2012ee} 
of $\sim$1.2 TeV on the gluino and first and second generation squark masses in these classes of models, with the result that typical measures of fine-tuning in these scenarios are soaring to the level of one part in $10^3$ or higher.  This has led to the claim that the constrained MSSM (cMSSM) paradigm is ``hardly tenable"~\cite{EuropeStudy} at this time.

Given this situation, it is imperative to explore the parameter space of supersymmetry in more detail, in order
to leave no corner unturned.  The p(henomenological)MSSM is particularly well-suited to this task.  In
this scenario, no specific theoretical prejudice is introduced at the GUT scale, or associated with a SUSY breaking mechanism, and a short list of experimentally motivated considerations reduce the MSSM to 19 (or 20) real,
weak-scale parameters, corresponding to a neutralino (or gravitino) Lightest Supersymmetric Particle 
(LSP)~\cite{Djouadi:1998di}.  This allows for the study of
LHC results in a more general fashion, and considers mass patterns and signatures which are not possible in the cMSSM and
other more conventional SUSY scenarios.  The pMSSM has thus garnered much attention in the 
literature~\cite{us,CahillRowley:2012cb,CahillRowley:2012rv,them}.  In particular, we have recently
shown~\cite{CahillRowley:2012cb,CahillRowley:2012rv} that the 7 TeV LHC SUSY searches allow for pMSSM models with light gluinos and 
squarks, and that a 126 GeV Higgs is easily accommodated in minimal supersymmetric models beyond the most constrained scenarios such as mSUGRA.  In addition, we found a handful of models in
a specific corner of parameter space with light stops/sbottoms and Higgsinos that allows for 1\% fine-tuning 
or better.

However, the LHC experiments have recently released a plethora of new search results which encompass 8 TeV 
updates of 7 TeV results, new signatures and new techniques.  Here, we re-examine the pMSSM in
light of this new data, and determine if our previous encouraging results still stack up.  In particular, we
investigate whether
the recently released 3rd generation LHC SUSY searches, involving top and bottom final states, exclude
our pMSSM models with light stops/sbottoms and low fine-tuning.  This provides the motivation for the present
work, where we examine all of the ATLAS SUSY related search results released before mid-September 2012;
this corresponds to 22 analyses with $\sim$5 fb$^{-1}$ at 7 TeV and $\sim$6 fb$^{-1}$ at 8 TeV.

The set of pMSSM models we consider were previously generated~\cite{CahillRowley:2012cb} by imposing the following set
of minimal assumptions on the R-parity conserving MSSM:  ($i$)  CP conservation, ($ii$) minimal flavor
violation at the electroweak scale, ($iii$) degeneracy of the first and second generations of sfermion
masses, ($iv$) the first two generations have negligible Yukawa couplings and A-terms.  In addition it
is assumed that the LSP is the neutralino\footnote{We have also generated
a large set of pMSSM models with a gravitino LSP, but these have some distinct collider signatures and
will be considered elsewhere. We do not consider sneutrino LSPs, though we note the existence of constraints from direct detection in the special case where the sneutrino makes up all of the dark matter.}.  We perform a random scan employing flat priors over the following parameter
ranges:  100 GeV $\leq m_{L_{1,3},e_{1,3}} \leq$ 4 TeV, 400 GeV $\leq m_{Q_1,u_1,d_1} \leq$ 4 TeV,
200 GeV $\leq m_{Q_3,u_3,d_3} \leq$ 4 TeV, 50 GeV $\leq |M_1| \leq$ 4 TeV, 100 GeV $\leq |M_2,\mu| \leq$ 4 TeV,
400 GeV $\leq |M_3| \leq$ 4 TeV, $|A_{\tau,t,b}|\leq$ 4 TeV, 100 GeV $\leq M_A \leq$ 4 TeV, and
1 $\leq \tan\beta \leq$ 60.  The limits of the scan ranges were chosen to obtain models that would test the search capabilities of the LHC at 7, 8, and 14 TeV. We generated 3 million points\footnote{We use the terms ``models'' and ``points'' interchangeably to refer to sets of pMSSM parameters and their associated spectra.} and subjected them to the global data set,
including, ($i$) theoretical considerations of no tachyons or color/charge breaking minima as well as
stable vacua, ($ii$)
precision electroweak data ($\rho$-parameter, invisible width of the $Z$ and the $W$ boson mass), 
($iii$) heavy flavor physics ($b\to s\gamma$, $B\to\tau\nu$, and $B_s\to\mu\mu$), 
($iv$) collider searches (LEP and Tevatron direct Higgs and SUSY searches, and LHC stable particle
searches), and ($v$) astroparticle physics constraints (dark matter direct detection and the WMAP relic
density, which we employ only as an upper bound).  We note that because of the relic density constraint in ($v$), many of our models have nearly pure wino or Higgsino LSPs. Roughly 225k models survived these constraints and
form our core pMSSM model sample~\cite{CahillRowley:2012cb}.  Of these, approximately 45k models contain a lightest
Higgs boson mass in the range $126\pm 3$ GeV and form what we will refer to as our ``Higgs subset"~\cite{CahillRowley:2012rv}.
For further details, please see~\cite{CahillRowley:2012cb,CahillRowley:2012rv}.

In what follows, we will find that even with more energy and more searches at the LHC, the pMSSM retains
the favorable properties listed above.  Specifically, we will see that light ($\lsim 600-700$ GeV) 
1st/2nd generation squarks and gluinos are consistent with the data, light ($\lsim 400$ GeV) stops/sbottoms 
can be present, and most of our low fine-tuning models remain viable.  In addition we will see that increased
luminosity at 8 TeV results in only marginal improvements in the reach of the standard ``vanilla'' searches.  We will examine the causes and
consequences of these results.  In particular, we will see that 
light Higgsinos can be kinematically accessible at the ILC, and that the next generation of dark
matter direct detection experiments have sensitivity to the low fine-tuning parameter region. In the
next section we describe the recent LHC SUSY analyses that we incorporate, as well as the necessary
modifications we implemented in the standard fast detection simulation PGS~\cite{PGS}.  In Section 3 we present our
results and their implications before concluding.

\section{LHC SUSY Searches}
\label{sec:analyses}

In this section we describe the LHC search channels employed in our study and the corresponding implications of our fast detector simulations. As in previous work~\cite{CahillRowley:2012cb}, we generally use the published limits on the numbers of events in each experimental search region, or compute the limits from the published numbers of expected background and measured events using the $CL_s$ method~\cite{Read:2002hq}.

\subsection{Vanilla, Multilepton, and 3rd Generation Searches}

We begin by presenting the LHC analyses considered for this study, where we closely follow the ATLAS search strategies. Generally, we have included every ATLAS SUSY search publically available as of September 2012 that is expected to significantly constrain the neutralino LSP pMSSM model set described above. An exception is the searches involving final-state taus, due to the current limitations of the fast detector simulator PGS~\cite{PGS} which we employed. In order to apply the ATLAS analyses, we have made numerous modifications to the PGS code, aiming to mimic the experimental environment as closely as possible. Details on our general strategy, and some of these changes, may be found in~\cite{CahillRowley:2012cb}. Here, we describe the additional modifications that were necessary to reproduce the ATLAS searches for this work.

First, we examine the ATLAS $\sim$5 fb$^{-1}$ 7 TeV jets plus MET~\cite{ATLAS:2012aa}, multi-jets~\cite{ATLAS:2012bb}, and one lepton~\cite{ATLAS:2012cc} searches. These were considered previously~\cite{CahillRowley:2012cb}, but remain relevant even after the 8 TeV data, as will be seen below. We will refer to these searches as the ``vanilla'' 7 TeV SUSY analyses, as opposed to the more specialized search channels, with signal regions involving third generation quarks or additional leptons.

Now that first and second generation squarks have been excluded at low mass in most standard scenarios, there has been much recent interest in searching explicitly for light third generation squarks. Such squarks would, of course, be particularly appealing for a supersymmetric resolution of the hierarchy problem. Several LHC analyses have sought evidence of direct stop or sbottom production by considering final states with $b$-quarks, both with and without leptons, and we have implemented many of them in this work. The various analyses target different squark mass ranges. As our scan ranges for the third generation squark mass parameters start at 200 GeV, we have not considered searches that are directed towards stops or sbottoms lighter than the top quark mass. Examining one such search~\cite{ATLAS:2012dd}, we have verified that it is very unlikely that a pMSSM model with a third generation squark above the top quark mass would be excluded, yet missed by more conventional searches for heavier third generation squarks. With this in mind, we have implemented all relevant ATLAS searches for third generation squarks heavier than the top quark~\cite{Aad:2012cz,ATLAS:mediumstop,:2012pq,:2012si,ATLAS:directsbottom,:2012ar}.
\begin{figure}
\centerline{\includegraphics[width=\textwidth]{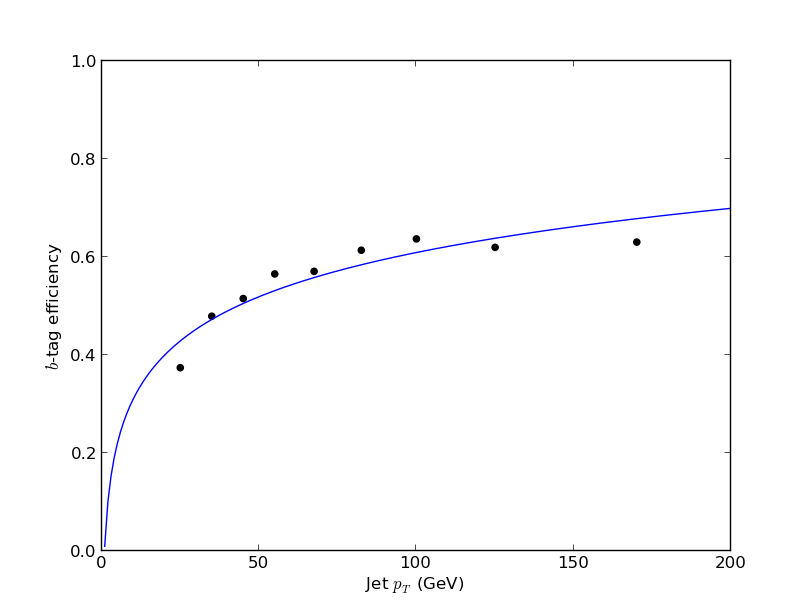}}
\caption{$b$-tagging efficiency as a function of $p_T$ for the \emph{MV1} algorithm. The points show ATLAS data obtained using a $t\bar{t}$ sample at an operating point where the algorithm's overall efficiency was 60\%. The curve is our fit to the ATLAS data. We use the fit function on true $b$-jets to implement $b$-tagging for signal regions using this algorithm and operating point. Similarly good fits are obtained for other algorithms and operating points.}
\label{fig:btag}
\end{figure}

The third generation searches use a variety of $b$-tagging algorithms at different efficiency points, reflecting both the rapid evolution of $b$-tagging methods and the higher $b$-tagging efficiencies required by recent searches involving multiple $b$-quarks. We have removed the original PGS $b$-tagging efficiency functions, which are based on the CDF Run II detector, and instead employ the built-in PGS ``truth'' jet type function for each jet, which outputs the type of the highest energy parton within 20 degrees of a jet. When testing whether an event passes the cuts for a given signal region, we take the true type of each jet in the event and apply an appropriate $b$-tagging efficiency function or light quark/gluon rejection factor. The efficiency functions and rejection factors vary for different signal regions and on the $b$-tagging algorithm chosen by ATLAS; we estimate these from fits to ATLAS data~\cite{btag}. As an example, Figure~\ref{fig:btag} shows the efficiency function we applied to true $b$-jets in order to reproduce the \emph{MV1} algorithm at an operating point that gave 60\% efficiency on a $t\bar{t}$ sample. By fitting the ATLAS data with a logarithmic function for each $b$-tagging algorithm at each operating efficiency used, we are able to match the experimental $b$-tagging procedure well.

There are also LHC searches using 7 TeV data with signal regions involving multiple leptons. These are directed at both colored sparticle production with subsequent decays through electroweak gauginos, as well as direct slepton/gaugino production. In principle, both of these channels are possible in the pMSSM, where the sparticle spectrum has no \emph{a priori} structure. Again, we do not include searches involving taus because PGS does not currently realistically reproduce tau efficiencies and fake rates at the LHC. However, all current ATLAS analyses involving electrons and muons are implemented~\cite{multilepton}.

In addition, searches have already been reported using the 8 TeV LHC data from 2012. In particular, ATLAS has performed analyses for final states with jets plus MET~\cite{ATLAS:2012ee}, many jets~\cite{ATLAS:2012ff}, one lepton with jets plus MET~\cite{ATLAS:2012gg}, and same-sign dileptons with jets plus MET~\cite{ATLAS:2012hh}, using $\sim$6 fb$^{-1}$ of 8 TeV data. Many of the signal regions in these analyses are quite similar to those of the corresponding 7 TeV searches, though some cuts are slightly tightened. As we will see in Section~\ref{results}, these small changes for the 8 TeV signal regions have the effect of reducing sensitivity to some models relative to the older 7 TeV searches. As in~\cite{CahillRowley:2012cb}, we also extrapolate the 8 TeV analyses to 25 fb$^{-1}$ of integrated luminosity, asking how many models would be excluded if the searches were repeated with the same cuts but more data. For this extrapolation, we scale the background and corresponding error in each signal region and calculate the signal event limit using the CL$_s$ method~\cite{Read:2002hq}, assuming that the observed number of events at 25 fb$^{-1}$ matches the expected background. Because of downward fluctuations in the observed numbers of events in many 8 TeV signal regions at 6 fb$^{-1}$, the estimated sensitivity with 25 fb$^{-1}$ can be comparable to, or even worse than, that obtained at 6 fb$^{-1}$. In fact, we will find that the fraction of models in our set that would be expected to be excluded with 25 fb$^{-1}$ of data at 8 TeV is not significantly higher than the fraction of models that is already excluded.

\subsection{Long-Lived Sparticles}

In addition to the standard searches using jets, leptons, and missing energy, searches for more exotic detector signals, particularly those associated with long-lived charged particles, provide important limits on supersymmetric models. For models with neutralino LSPs, the lightest chargino is frequently metastable when the LSP is a nearly pure wino eigenstate. In this case the next-to-LSP (NLSP) is usually a wino-like chargino, and the mass splitting between the two is typically only $\sim$160 MeV, leading to detector-scale decay lengths. The prevalence of nearly pure wino LSPs in our model set~\cite{CahillRowley:2012cb} results in a significant fraction ($\sim$4\%) of models in which the lightest chargino has an unboosted decay length over 7m (\textit{e.g.}, long enough to traverse the CMS muon spectrometer), and even more models ($\sim$25\%) in which the lightest chargino decays in the tracker, calorimeter, or muon chamber. In a few cases, sparticles other than the $\chi_1^\pm$ are metastable as a result of small mass splittings with the LSP. However, these cases are quite rare since they rely on accidental degeneracies that are uncommon due to the random nature of our parameter scan. As a result, only a tiny fraction (0.14\%) of our models have charged sparticles that are not charginos and have unboosted decay lengths beyond 7m. Our analysis is thus tailored to chargino NLSPs, although it is applicable to any uncolored metastable sparticle. Models with metastable colored sparticles require modifications to hadronization and decay routines within PYTHIA and will be considered in our study of the collider signatures of the pMSSM with a gravitino LSP~\cite{future}.

Metastable charginos produce several different signatures depending on the location of their decay vertex. Charginos decaying inside the tracker produce an extremely soft pion, resulting in a disappearing track which is the focus of a dedicated ATLAS search~\cite{ATLAS:2012ii}. Charginos decaying outside of the tracker will be detected by searches for highly ionizing particles, using either the inner detector alone (if the chargino decays before leaving the muon spectrometer) or both the inner detector and muon chamber. Both ATLAS and CMS search for highly ionizing particles; we choose to apply the CMS search~\cite{Chatrchyan:2012sp} to our model set as the variety of signal channels and efficiencies presented are helpful for validation of our simulation.

Metastable charginos require careful treatment in several stages of our analysis. As described in~\cite{us}, we employ analytic formulae to calculate chargino decays for models in which the lightest chargino and LSP masses are split by less than 1 GeV. We have also found it necessary to modify the PGS package. Within PGS, we have altered, or added, functions reconstructing a metastable charged particles's momentum, velocity, and calorimeter response, and have modified the output to include the location of any decay vertex, the presence or absence of a muon spectrometer track, the tracker and calorimeter isolation sums, and the particle velocity. Our analysis routine uses the particle velocity to compute the ionization energy loss in the tracker before applying cuts specific to the individual search. These modifications are described in detail below.

The most sensitive searches for very long-lived charginos rely heavily on the muon system. We therefore introduce several modifications to the standard PGS treatment of the muon spectrometer response to a heavy metastable charged particle. First, we have implemented a special momentum resolution function for particles reconstructed in the muon spectrometer with $p_{_T}$ $>$ 100 GeV. This is necessary because the standard spectrometer momentum resolution function in PGS is simulated by the momentum resolution of the inner tracker, which is outperformed by the muon spectrometer at high $p_{_T}$. Our muon spectrometer momentum resolution function is a copy of that used to simulate the inner detector momentum resolution, but employs input parameters mimicking the ATLAS muon spectrometer. (The difference in resolution between the ATLAS and CMS muon systems is not important for our purposes.) 
The main motivation for including a dedicated muon resolution function is to decrease the extraneous MET resulting from mismeasurement of long-lived sparticles that frequently have a very high $p_{_T}$. Second, we have strengthened the requirements for track reconstruction within the muon spectrometer (which was previously only a pseudorapidity cut). A metastable charged particle can fail reconstruction in the muon spectrometer by decaying before it produces enough hits for reconstruction. Since the requirements for reconstruction depend on the detailed geometry of the muon spectrometer, we conservatively require muon candidates to decay outside of the muon spectrometer ($r \geq 7$ m in the rapidity region of interest). Additionally, slow-moving particles may produce delayed hits in the muon spectrometer that are not associated with the primary vertex. We therefore subject the track reconstruction to the velocity-dependent efficiency given in~\cite{exo10004}:
\begin{equation}
f_{acc} = \left\{ \begin{array}{ccc}
0 & \mathrm{for} & \beta\ < ~0.3  ,\\
2.5 \beta\ - 0.75 & \mathrm{for} &  0.3 < \beta\ \leq 0.7 ,\\
1 & \mathrm{for} & \beta\ > 0.7 ,\\
\end{array} \right
.\end{equation}       
where $f_{acc}$ is the fraction of stable charged particles with a given velocity and a rapidity within the muon spectrometer acceptance that are reconstructed by the muon system. Finally, timing measurements within the muon spectrometer provide a measurement of $\beta^{-1}$ by using time-of-flight. We simulate this measurement by smearing the truth-level value of $\beta^{-1}$ with a Gaussian with a width of 0.06, as given by~\cite{Chatrchyan:2012sp}. 

Although muons and muon-like particles typically deposit only a few GeV of energy in the calorimeters, very slow-moving particles ($\beta\ \lesssim 0.1$) can deposit a significantly larger amount of energy. Since few metastable charginos are produced with such a low velocity, however, we simply neglect the calorimeters and use the muon track to correct the missing energy measurement. While this approach slightly overestimates the amount of missing energy in events with slow-moving metastable particles which fail to produce a muon track, the effect on our results is insignificant.

The CMS search for heavy stable charged particles uses a modified version of the Smirnov-Cramer-von Mises discriminant, denoted by $I_{as}$, to estimate the compatibility of the observed charge deposition for a given track and that expected for a particle near the minimum of ionization. The discriminant is defined by the following formula:
\begin{equation}
I_{as} = \frac{3}{J} \times { \frac{1}{12J} + \sum\limits_{i=1}^J [P_i \times (P_i-\frac{2i-1}{2J})^2]},
\end{equation}
where $J$ is the number of hits and $P_i$ is the sorted probability for the energy deposited by a minimum-ionizing particle to be less than or equal to the energy deposited in the $i^{th}$ hit. Our analysis code uses our modified PGS output to calculate $I_{as}$ in the following manner. First, the \textit{most probable} ionization is calculated using an experimental parametrization:
\begingroup
\begin{equation}
\left(\frac{dE}{dx}\right)_{_{mp}} = K \frac{m^2}{p^2} + C,
\end{equation}
\endgroup
with $K = 2.559 ~ \frac{\mathrm{MeV} \mathrm{c}^2}{\mathrm{cm}}$ and $C = 2.772 ~\frac{\mathrm{MeV}}{\mathrm{cm}}$~\cite{Chatrchyan:2012sp}. This expression agrees closely with the Bethe-Bloch formula in the range $0.4 \leq \beta \leq 0.9$; for values of $\beta$ below (above) this range, the precise details can be neglected because the particle is clearly highly ionizing (minimum ionizing). Second, we assume that the particle has 16 hits in the CMS tracker~\cite{CMS:2012aa}, and choose an energy deposition $(\frac{dE}{dx})_{_{hit}}$ for each hit from a Gaussian of width $0.559 ~\frac{\mathrm{MeV}}{\mathrm{cm}}$ centered at $(\frac{dE}{dx})_{_{mp}}$. (The width is determined by a Gaussian fit to the ionization energy distribution for minimum bias events in Figure 4 of~\cite{Khachatryan:2010pw}.) We model the saturation of the CMS pixel detector with a threshold of $9.436~\frac{\mathrm{MeV}}{\mathrm{cm}}$ by assigning the threshold value to any hit above threshold. Third, for each hit we calculate the probability that a minimum-ionizing particle would deposit a smaller or equivalent amount of energy by integrating the probability density function (PDF) for its ionization energy deposition from zero to $(\frac{dE}{dx})_{_{hit}}$.{\footnote {We take the PDF from Figure 4 of ~\cite{Khachatryan:2010pw}. While this distribution is generated from pixel detector hits for a sample of minimum bias events at $\sqrt{s} = 0.9$ TeV, most of these hits come from pions near the minimum of ionization. As a cross-check, the PDF for minimum ionizing particles determined in a test run~\cite{Tinti:2011rk} is similar to the one we use. We also expect that the PDFs for the silicon strip tracker and pixel detector are equivalent within our desired accuracy.}} Finally, we sort the individual probabilities and sum them according to the expression given above.

The CMS search for heavy stable charged particles (HSCPs)~\cite{Chatrchyan:2012sp} requires an inner detector track with 11 hits in the silicon tracker. We approximate this high number of required hits by simply requiring a HSCP candidate to decay outside of the silicon tracker, r $\geq$ 1.1 m~\cite{Chatrchyan:2008aa}. We simulate the preselection process by imposing cuts on MET, candidate pseudorapidity, and calorimeter and tracker isolation sums. For the tracker+time-of-flight analysis, we additionally require the presence of a track in the muon spectrometer. The final selection is divided into search regions, defined by cuts on the values of $p_{_T}$, $I_{as}$, the reconstructed mass, and (for the tracker+time-of-flight analysis) $\beta$. We simulate the reconstructed mass cut by applying a given search region only to candidates with a truth-level mass at least 1 standard deviation above the cut value. This implies that an insignificant fraction of signal events will be removed by the mass cut. We note that all of our stable particles pass the mass cut for at least one search region, and even low-mass search regions have very high sensitivity. We validated this analysis by comparing our simulation with two benchmarks: Pair-produced stable staus with masses of 100 GeV and 308 GeV. The results of our validation are shown in Tables~\ref{CMS_tkonly} and~\ref{CMS_tktof}. Comparing our results with those of CMS shows that we reproduce the production cross section for both benchmarks. The acceptance of the tracker+time-of-flight analysis is also reproduced quite nicely, while our acceptance for the tracker-only analysis (particularly for the 100 GeV stau) is significantly higher than that predicted by CMS. Since the tracker-only analysis is only relevant for charginos decaying in the calorimeter or muon spectrometer, the tracker+time-of-flight analysis is considerably more important for our model set. We note that the discrepancy seen for the light stau model is most likely because the light staus tend to have $\beta \sim1$, making them difficult to distinguish from minimum-ionizing particles and introducing a strong sensitivity to the details of ionization energy deposition. Fortunately, models with very light stable charginos are excluded by the LHC, the Tevatron and LEP (and therefore are excluded during model generation). The accuracy of our treatment is significantly better where it is most relevant, namely for heavier charginos with masses near the the exclusion boundary.

The ATLAS disappearing track search~\cite{ATLAS:2012ii} looks for charginos which are reconstructed in the silicon tracker (SCT), but not in the outer transition radiation tracker (TRT). We therefore require candidates to decay between the outer edge of the SCT (520 mm) and the inner edge of the outer TRT (863 mm). The high number of hits required in the SCT and low number required in the outer TRT support these sharp cutoffs. We approximate the track isolation requirement (no tracks with $p_{_T} > 0.4$ GeV within $\Delta R \leq 0.1$) by using $\Delta R \leq 0.25$; a negligible difference between the two requirements was observed during comparison with benchmarks. We then require candidates to be the highest $p_{_T}$ isolated track in the event and impose cuts on the standard physics objects (leptons, jets, and MET). We validated our analysis with the ATLAS benchmarks (100 and 200 GeV charginos with 1 ns lifetimes); the results are presented in Table~\ref{ATLAS_distrk}. Our pair-production cross sections and signal efficiencies for both models are in agreement with the ATLAS values, indicating that our geometric acceptance cuts and isolation requirements accurately reproduce the ATLAS analysis.

\begin{table}
\centering
\begin{tabular}{|l|c|c|c|c|} \hline\hline
Benchmark    &  $\sigma$ (us)  &  A (us) & $\sigma$ (CMS)  & A (CMS) \\ \hline
100 GeV $\tilde{\tau}_1$     &  41 fb &  20\% & 38 fb & 11\% \\
308 GeV $\tilde{\tau}_1$     &  0.42 fb &  59\% & 0.35 fb & 39\% \\
\hline\hline
\end{tabular}
\caption{Validation of our simulation of the CMS tracker-only HSCP search, comparing our results for the production cross section and acceptance with that of CMS.}
\label{CMS_tkonly}
\end{table}

\begin{table}
\centering
\begin{tabular}{|l|c|c|c|c|} \hline\hline
Benchmark    &  $\sigma$ (us)  &  A (us) & $\sigma$ (CMS)  & A (CMS) \\ \hline
100 GeV $\tilde{\tau}_1$     &  41 fb &  21\% & 38 fb & 19\% \\
308 GeV $\tilde{\tau}_1$     &  0.42 fb &  63\% & 0.35 fb & 55\% \\
\hline\hline
\end{tabular}
\caption{Validation of our simulation of the CMS tracker+time-of-flight HSCP search, comparing our results for the production cross section and acceptance with that of CMS.}
\label{CMS_tktof}
\end{table}

\begin{table}
\centering
\begin{tabular}{|l|c|c|c|c|} \hline\hline
Benchmark    &  $\sigma$ (us)  &  A (us) & $\sigma$ (ATLAS)  & A (ATLAS) \\ \hline
100 GeV $\tilde{\chi}^\pm_1$     &  13100 fb &  0.070 \% & 14400 fb & 0.066 \% \\
200 GeV $\tilde{\chi}^\pm_1$    &  819 fb &  0.11 \% & 808 fb & 0.12 \% \\
\hline\hline
\end{tabular}
\caption{Validation of our simulation of the ATLAS disappearing track search, comparing our results for the production cross section and acceptance with that of ATLAS.}
\label{ATLAS_distrk}
\end{table}

\section{Results and Implications}
\label{resimp}

In this section we will discuss the results of the various SUSY searches and their implications for the parameter space of our pMSSM model set with a neutralino LSP.

\subsection{Search Results}
\label{results}

The first question we address is how well these searches perform, both individually and when combined, in covering the set of pMSSM models that we generated. We 
begin by considering the ``vanilla,'' generalized jets plus MET, searches at 7 and 8 TeV. Tables~\ref{7TeV-vanilla} and~\ref{8TeV-vanilla} show the fraction of our model 
set excluded by these two suites  
of searches, individually and when they are combined. Note that results are shown for both the full model set and for the subset of models satisfying m$_h$ = $126\pm 3$ GeV, henceforth referred to as the ``Higgs subset''. 
There are several things to observe about these results: ($i$) Individual models are more than likely to be excluded by more than one of these searches 
as is clear when we compare the combined results to those for the individual searches themselves. ($ii$) The 2-6 jets plus MET analysis is by far dominant and provides 
most of the model coverage at both center of mass energies. This is different from the ATLAS results for their corresponding cMSSM SUSY searches \cite{ATLAS:2012ee,ATLAS:2012aa,ATLAS:2012bb,ATLAS:2012cc,ATLAS:2012ff,ATLAS:2012gg,ATLAS:2012hh};
there the coverage provided by the 1-lepton plus jets search channel was found to be relatively comparable to that of the 2-6 jets plus MET channel~\cite{Paige}. There are several reasons why the leptonic searches are degraded in the pMSSM case, the most important being the lack of light 
binos in the spectrum. As we have discussed earlier~\cite{CahillRowley:2012cb}, most of the LSPs in our pMSSM model set are either nearly pure wino or Higgsino due to our requirement that the LSP relic density be below the WMAP measurement, and these models tend to have 2 or 3 nearly degenerate states at the bottom of their spectra. On the other hand, in the cMSSM the LSP is quite commonly bino-like (with no other gaugino states very close 
in mass) with a somewhat heavier wino. This then allows for wino decays through gauge bosons which subsequently produce leptons with sufficient $p_{_T}$ to pass analysis cuts in 
the cMSSM, whereas the corresponding leptons in the pMSSM will be too soft. In the subset of pMSSM models with relatively light bino-like neutralinos we indeed 
find an improved efficiency for model coverage by the leptonic channels. We might also expect the performance of leptonic channels to improve significantly with increasing luminosity, which will increase the sensitivity of these channels to final states with low branching fractions, \textit{e.g.}~leptons produced in cascade decays.
($iii$) We note that employing the Higgs mass constraint does not significantly alter the fractions of models excluded by the various SUSY search channels in a 
qualitative manner, although there is some degradation ($\sim10$\%) observed in the coverage provided by each of the searches individually as well as in the combination. ($iv$) Comparing
the results obtained at 7 TeV with $\sim$5 fb$^{-1}$ and those at 8 TeV with $\sim$6 fb$^{-1}$, we see that the 8 TeV data increases the {\it total} amount of model coverage 
by $\sim$25\% and that the model coverage provided by the searches individually improves. This implies that more models are being excluded by multiple 
searches at 8 TeV than are at 7 TeV. 

Interestingly, a small number of models ($\sim$1.6k) are found to be excluded by the 7 TeV searches but {\it not} by 
the corresponding channels at 8 TeV. This is likely due to the increased jet $E_{_T}$ and MET requirements of the 8 TeV searches that are now allowing models with somewhat 
degenerate spectra to be missed. There is a valuable lesson to be drawn from this result. As the LHC energy is further increased, it will be likely that some reasonable 
fraction of pMSSM model points with spectrum degeneracies will be bypassed by the analyses due to the continual strengthening of analysis cuts. For more complete model 
coverage it is then clearly useful and necessary to combine analyses that are performed at different center of mass energies, or to perform targeted searches for compressed spectra~\cite{LeCompte:2011cn}. In our case of interest, however, combining the 7 and 8 TeV generalized analyses we find that the total model coverage increases by only $\sim$1\%.

\begin{table}
\centering
\begin{tabular}{|l|l|c|c|} \hline\hline
Search  &   Reference          &   Full Model Set  & Higgs Subset \\ \hline
2-6 jets & ATLAS-CONF-2012-033  &  21.04\%  &  18.53\%  \\
multijets & ATLAS-CONF-2012-037  &  1.61\%   & 1.34\%  \\
1-lepton  & ATLAS-CONF-2012-041  &  3.16\%  & 2.80\%  \\
Total  &      &  21.19\%  & 18.64\%  \\
\hline\hline
\end{tabular}
\caption{Fraction of our pMSSM models with neutralino LSPs excluded (in per cent) by the general ``vanilla'' MET ATLAS searches at the 7 TeV LHC with 4.7 fb$^{-1}$ of 
integrated luminosity for both the full model set as well as for the ``Higgs Subset'' satisfying the Higgs mass constraint, $m_h=126\pm 3$ GeV.}
\label{7TeV-vanilla}
\end{table}

\begin{table}
\centering
\begin{tabular}{|l|l|c|c|} \hline\hline
Search  &   Reference          &   Full Model Set  &  Higgs Subset \\ \hline
2-6 jets   &   ATLAS-CONF-2012-109     &  26.51\%  &  23.82\%  \\
multijets   &  ATLAS-CONF-2012-103  &  3.31\%   & 2.84\%  \\
1-lepton     &  ATLAS-CONF-2012-104  &  3.30\%   & 3.07\%  \\
SS dileptons &  ATLAS-CONF-2012-105  & 4.88\%   & 4.50\%  \\
Total        &     & 26.90\%  & 24.16\%  \\
\hline\hline
\end{tabular}
\caption{Same as Table~\ref{7TeV-vanilla} but now for the 8 TeV, 5.8 fb$^{-1}$ ATLAS searches.}
\label{8TeV-vanilla}
\end{table}

Given these results, we can ask how well these generalized jet plus MET searches at 8 TeV will perform when an expected luminosity of 25 fb$^{-1}$ is reached by the end of the 2012 run. To do 
this kind of extrapolation we need to make several assumptions: (a) the ATLAS analyses are unchanged from their present form in terms of cuts, \etc. (although 
these are likely to be strengthened); (b) we can simply scale the SM backgrounds found by ATLAS in the lower luminosity data to the case of 25 fb$^{-1}$ 
and (c) the number of events observed in each analysis channel is exactly matched to the corresponding anticipated backgrounds.{\footnote {If ATLAS is ``lucky'' and observes fewer events than anticipated (as frequently happens) then stronger constraints will naturally result from the relevant analysis.}}  
With this set of conservative assumptions we obtain the results shown in Table~\ref{8TeV-extrap}. Here we see that whereas for the two leptonic searches pMSSM model coverages 
are increased relative to the 
lower luminosity results, those for the 2-6 jets plus MET search are actually {\it degraded} whereas those for the multijet final state are essentially unchanged. The 
reason for this is the result of our assumption (c) above and the fact that ATLAS saw {\it fewer} events than expected in many of their 2-6 jets plus MET search regions with 5.8 fb$^{-1}$. Clearly 
the exact coverage will be quite sensitive to the actual observed signal vs. background ratio and this situation will need revisiting subsequent to the publication of the 
complete ATLAS 25 fb$^{-1}$ results sometime next year. Overall, given our assumptions, it seems that higher luminosities may not cover too much more of the parameter 
space. What is really needed for more coverage is to go to higher energies, $\sqrt s=13-14$ TeV.

\begin{table}
\centering
\begin{tabular}{|l|l|c|c|} \hline\hline
Search  &   Reference          &   Full Model Set  &  Higgs Subset \\ \hline
2-6 jets     & ATLAS-CONF-2012-109  &  25.27\% &  22.68\%  \\
multijets    & ATLAS-CONF-2012-103  & 3.31\%   & 2.84\%  \\
1-lepton     & ATLAS-CONF-2012-104 & 3.83\%   & 3.57\%  \\
SS dileptons & ATLAS-CONF-2012-105  & 7.45\%   & 6.96\%  \\
Total     &   &  26.05\%  & 23.49\%  \\
\hline\hline
\end{tabular}
\caption{Same as Table~\ref{8TeV-vanilla} but now extrapolated to 25 fb$^{-1}$ of integrated luminosity at 8 TeV.}
\label{8TeV-extrap}
\end{table}

We now turn our attention to the large set of Heavy Flavor (HF) and Multilepton (ML) searches which, to date, have only been reported by ATLAS for the $\sqrt s=7$ TeV data set, mostly with 
4.7 fb$^{-1}$ of integrated luminosity. As noted above, these searches are especially important as we would expect natural SUSY models to have relatively light stops/sbottoms 
as well as some light Higgsino-like (and possibly wino-like) gauginos. To date, 11 of these HF/ML searches, as shown in Table~\ref{HF-ML}, have been reported by ATLAS. As noted above, in order to analyze these HF 
channels, it was necessary to modify the b-tagging routines employed by PGS~\cite{PGS} to more accurately reflect those of the ATLAS detector~\cite{btag}. The results shown in 
Table~\ref{HF-ML} display the pMSSM coverage for each separate analysis, as well as for the two combinations of the set of HF and ML analyses. Here, again, the results are shown for both the full model set as well as for the subset of models with the Higgs mass constraint imposed. As in the case of the generalized MET searches, we see that applying 
the Higgs mass cut causes some 
small degradation ($\sim$10-15\%) in the sensitivity of these analyses as well as in the combinations. Clearly some of these searches will be more powerful at probing the pMSSM than are others. It is 
no surprise that the coverage is rather poor for the very light stop search (since all our stops are required to be more massive than the top quark in the model generation 
process), as well as for the direct stop in the natural GMSB search. Also, since the cross sections for direct gaugino production are rather small and, in most of 
our models, these states are quite degenerate due to their being relatively pure wino/Higgsino eigenstates, the direct gaugino search is also generally not very effective at probing our model set. Other 
searches, such as those for heavy stops, direct sbottoms, and 1-2 leptons plus jets, perform significantly better and provide reasonable model coverage. Since these searches 
are relatively orthogonal to the generalized jet plus MET searches discussed above, we would expect that HF/ML searches will add significantly to the overall pMSSM model 
coverage and this is indeed the case. In particular, we find that 0.96\%, 0.08\%, and 1.03\% of models are excluded by heavy flavor searches, multilepton searches, and their combination (respectively), but are not covered by the 7 and 8 TeV ``vanilla'' searches.

\begin{table}
\centering
\begin{tabular}{|l|l|c|c|} \hline\hline
Search  &   Reference          &   Full Model Set  & Higgs Subset \\ \hline
Gluino $\to$ Stop/Sbottom   &   1207.4686               &  4.92\%   &  4.54\%  \\
Very Light Stop  &    ATLAS-CONF-2012-059               &  $<$0.01\%  &   $<$0.01\%    \\
Medium Stop  &   ATLAS-CONF-2012-071                &  0.32\%   &  0.24\%  \\
Heavy Stop (0l)  &  1208.1447                 &  3.66\%   &  3.15\%  \\
Heavy Stop (1l)   &  1208.2590                &  1.94\%   &  1.69\%  \\
GMSB Direct Stop   &   1204.6736               &  $<$0.01\%  &   0\%    \\
Direct Sbottom  &    ATLAS-CONF-2012-106               &  2.47\%   &  2.20\%  \\
3 leptons  &   ATLAS-CONF-2012-108              &  1.05\%   &  0.92\%  \\
1-2 leptons  &    1208.4688               &  4.11\%   &  3.61\%  \\
Direct slepton/gaugino (2l)  &   1208.2884                 &  0.11\%   &  0.10\%  \\
Direct gaugino (3l)  &   1208.3144                &  0.33\%   &  0.27\%  \\
                 &         &       & \\
HF Total    &     &  7.26\%  & 6.48\%    \\
ML Total     &    &  4.29\%   & 3.78\%   \\
\hline\hline
\end{tabular}
\caption{Same as Table~\ref{7TeV-vanilla} but now for the HF and ML searches.}
\label{HF-ML}
\end{table}

In addition to searches with MET, those looking for heavy stable charged particles (HSCP) or disappearing tracks (DT) will be quite important for our pMSSM model set due to the 
rather large number of long-lived charginos present in these models. Such states will be either directly produced or may be present in low-MET cascades. In the latter 
case we can access long-lived charginos with masses beyond those accessible via direct production. Again the relatively frequent appearance of long-lived charginos in our model sample is due to the rather pure wino- and Higgsino-like nature of most LSPs and the relative degeneracies of the charginos and the LSP which occur in such cases. Table~\ref{Non-MET} 
shows the sensitivity of these searches for the full model set and for the Higgs subset. Interestingly, unlike the other searches that were discussed above, that 
for HSCP performs slightly better in the subsample of models where the Higgs mass constraint is applied.

\begin{table}
\centering
\begin{tabular}{|l|l|c|c|} \hline\hline
Search     &  Reference   &   Full Model Set  & Higgs Subset \\ \hline
HSCP       &  1205.0272   &  4.03\% &  4.14\%  \\
Dis. Tracks  & ATLAS-CONF-2012-111  &  2.59\%   & 2.21\%  \\
$B_s \to \mu^+\mu^-$  & ~\cite{atlasbs, atlasbs2, cmsbs, lhcbbs}   &  2.70\%   & 5.58\%  \\
$A/H\to \tau^+\tau^-$  &  1202.4083  & 0.07\%   & 0.03\%  \\
     &         &         &        \\
All Searches  & & 33.89\%  & 33.45\%  \\
\hline\hline
\end{tabular}
\caption{Same as Table~\ref{7TeV-vanilla} but now for the non-MET searches. The corresponding combined results obtained from all searches is also shown.}
\label{Non-MET}
\end{table}

Finally, before we combine the results from all these searches to extract the total pMSSM coverage at the LHC, we must recall the significant contributions arising from both the 
searches for $B_s\to \mu^+\mu^-$~\cite{atlasbs, atlasbs2, cmsbs, lhcbbs} as well as for $A,H\to \tau^+\tau^-$~\cite{Chatrchyan:2012vp} which were discussed in detail in our earlier work~\cite{CahillRowley:2012cb}. The former, in particular, is relatively 
powerful, and can allow us to access model points with heavier gluino and/or first and second generation squark masses than do the present MET searches. These results are reproduced in Table~\ref{Non-MET} where we see that for the subset of models where the Higgs mass constraint has been enforced the $B_s \to \mu^+\mu^-$ is significantly more 
effective.  

Combining all these various searches (from Tables~\ref{7TeV-vanilla},~\ref{8TeV-vanilla},~\ref{HF-ML} and~\ref{Non-MET}) together without (with) the Higgs mass constraint imposed, we see from Table~\ref{Non-MET} that 33.89 (33.45)\% of our pMSSM models are 
now excluded by the combined data from the LHC. Note that when all the searches are combined, the {\it overall} effect of the Higgs mass constraint is very weak. To make a 
significant improvement in these numbers the complete search results for 8 TeV will need to be included. Of course, going to $\sqrt s \approx 14$ TeV will provide a major step in
model coverage.

\subsection{Implications}
\label{implications}

\begin{figure}
\centering
\subfloat{
     \begin{overpic}[height=3.5in]{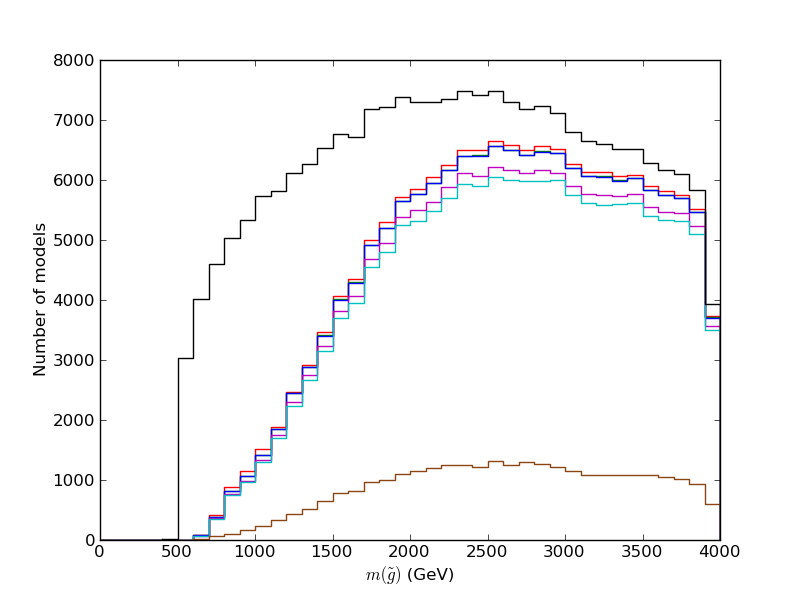}
     \end{overpic}
     } \\
\subfloat{
     \begin{overpic}[height=3.5in]{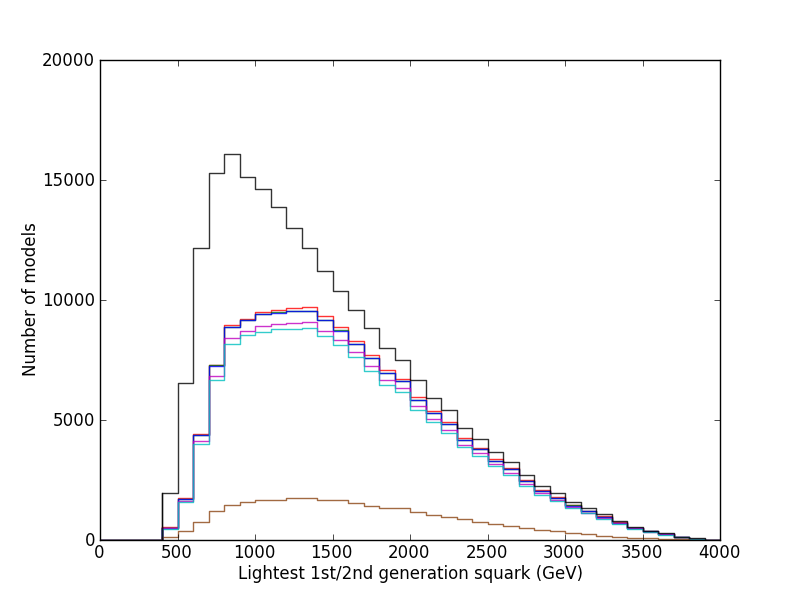}
     \end{overpic}
     }
\caption{Histograms showing the distribution of gluino (upper panel) and lightest 1st/2nd generation squark (lower panel) masses in our pMSSM model set with a neutralino LSP. The effect of sequentially applying LHC searches on these distributions is shown as a series of colored lines in the following order from top to bottom: Full Model Set (black), 7 and 8 TeV Vanilla Searches (red), Heavy Flavor (green), Multileptons (blue), HSCPs and Disappearing Tracks (magenta), $B_s \to \mu^+\mu^-$ and $H/A\to \tau^+\tau^-$ (cyan), and m$_h$ = $126\pm 3$ GeV (brown). The lower bin value at the right edge of the gluino mass distribution is due to our upper limit of 4 TeV for $M_3$, combined with radiative corrections which tend to decrease the gluino mass.}
\label{fig:SGhist}
\end{figure}

\begin{figure}
\centering
\subfloat{
     \begin{overpic}[height=3.5in]{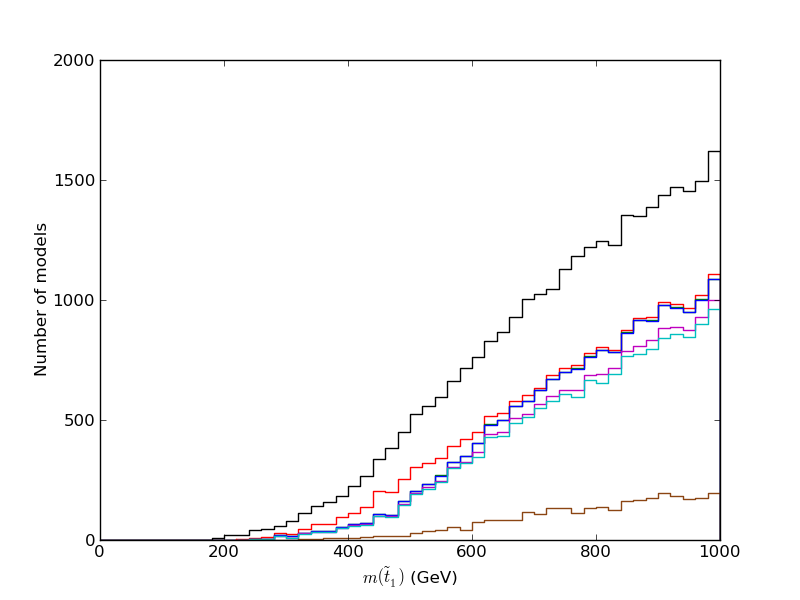}
     \end{overpic}
     } \\
\subfloat{
     \begin{overpic}[height=3.5in]{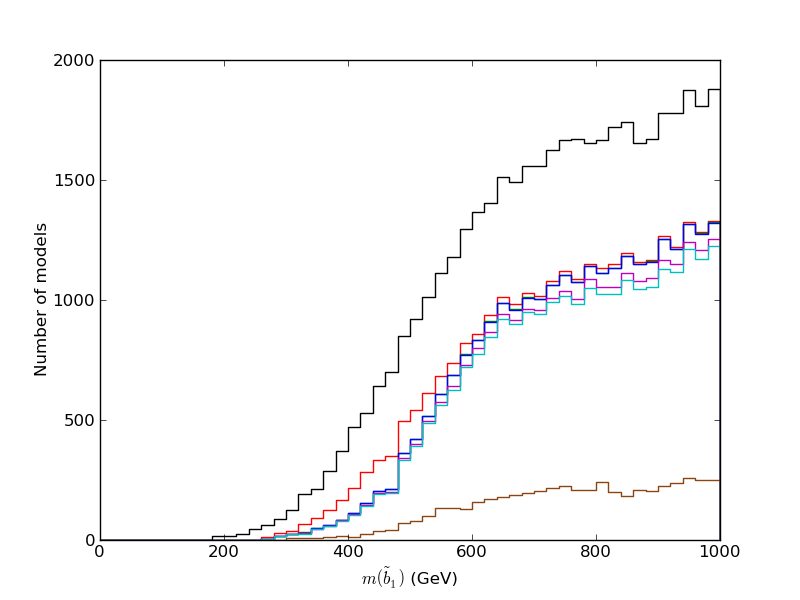}
     \end{overpic}
     }
\vspace*{0.5cm}
\caption{Histograms showing the distribution of the lightest stop (upper panel) and lightest sbottom (lower panel) masses in our pMSSM model set with a neutralino LSP. The line colors are as in Figure~\ref{fig:SGhist}.}
\label{fig:g3hist}
\end{figure}

\begin{figure}
\centering
\subfloat{
     \begin{overpic}[height=3.5in]{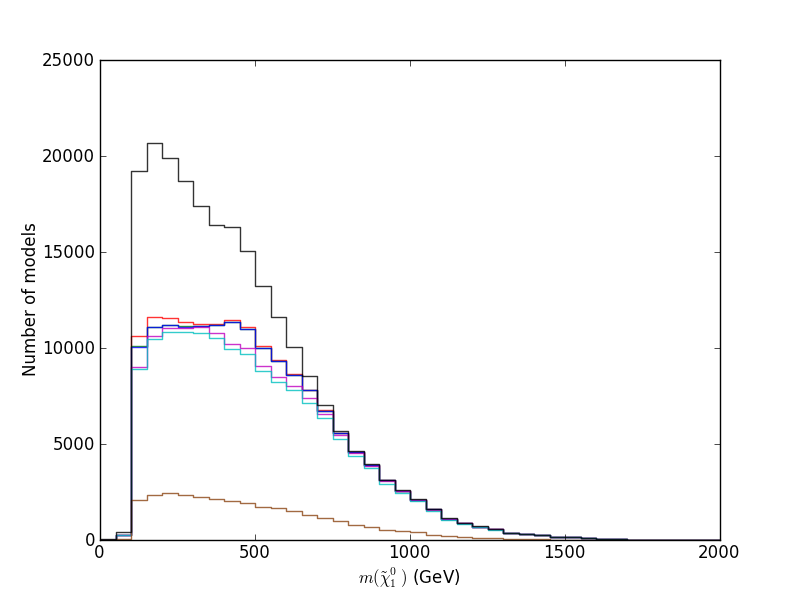}
     \end{overpic}
     } \\
\subfloat{
     \begin{overpic}[height=3.5in]{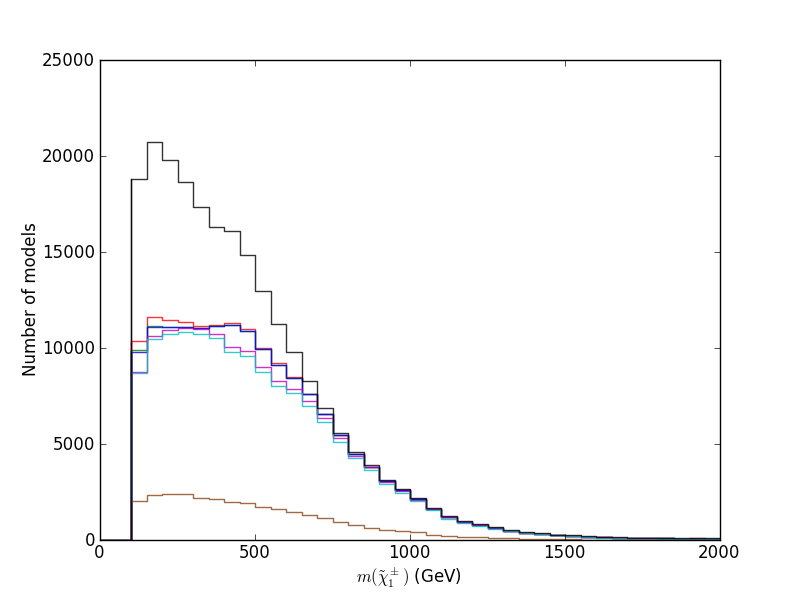}
     \end{overpic}
     }
\vspace*{0.5cm}
\caption{Histograms showing the distribution of the lightest neutralino (upper panel) and lightest chargino (lower panel) masses in our pMSSM model set with a neutralino LSP. The line colors are as in Figure~\ref{fig:SGhist}.}
\label{fig:EWGhist}
\end{figure}

We now turn to a discussion of some of the implications of these various searches. The first question we address is how the distributions of the various sparticle masses are influenced by the set of negative SUSY search results obtained so far at the LHC. Furthermore, we would also like to know if some mass ranges in 
the pMSSM are now entirely excluded by the LHC searches. These questions are answered by examining the nested set of histograms presented for the various sparticle types 
in Figures~\ref{fig:SGhist}-\ref{fig:EWGhist}.  

In these figures we see the distribution for the masses of various sparticles for models that remain valid after evading a series of LHC experimental searches, with each line indicating the effect of applying an additional search in the following order: 7 and 8 TeV Vanilla Searches, Heavy Flavor, Multileptons, HSCPs and Disappearing Tracks, $B_s \to \mu^+\mu^-$ and $H/A\to \tau^+\tau^-$, and m$_h$ = $126\pm 3$ GeV.

Let us first consider the case of the gluino. In Figure~\ref{fig:SGhist} we see, as expected, that the 7 and 8 TeV general MET searches take the largest bite out of the gluino mass 
distribution, particularly at the low mass end. The other searches tend to remove model points corresponding to heavier gluino masses which are not directly kinematically accessible. Interestingly, and we will discuss this further below, we see that a sizable fraction of models still exist with gluino masses below $\sim$700 GeV. In the 
case of the lightest first/second generation squarks, while the 7 and 8 TeV general MET searches remove many models from the distribution, they are most effective below 
$\sim$900 GeV while at larger masses the constraints are indirect. The HSCP searches are seen to be more effective in this case particularly in the intermediate mass region 
$\sim$1.0-1.5 TeV. We again see that models with somewhat lighter squarks with masses just above $\sim$600 GeV still remain viable.

Lighter stops/sbottoms are a common prediction in natural models with low fine-tuning, so it is particularly important to examine the sensitivity of the HF searches to these states. 
In Figure~\ref{fig:g3hist} we see the spectra for both light stops and sbottoms in the most interesting mass region below 1 TeV. In addition to the expected large impact of the 7 and 8 TeV 
general MET searches we see, particularly in the mass range 
below $\sim$ 650 GeV, that the HF searches are making a significant dent in the number of surviving models. However, the number of models with light stops or sbottoms 
below $\sim$ 400 GeV still remains significant. When the Higgs mass constraint is included there is seen to be a serious reduction in the number of remaining models with lighter 
stops (though as mentioned in Section~\ref{results}, the overall \emph{efficiency} of the LHC searches is not significantly altered). This is not too surprising 
as heavy stops, and/or large stop mixing, is required in the pMSSM to obtain Higgs masses in the required $126 \pm 3$ GeV range~\cite{CahillRowley:2012rv}. We note, however, that since the Higgs mass requirement 
does not directly affect the existence of lighter sbottoms, a significant number of models remain with relatively low sbottom masses. In fact, after applying the Higgs mass constraint 
the sbottoms are generally found to be lighter than the stops. The fact that relatively light stops/sbottoms remain in the spectra of the surviving models is not too surprising as 
the standard stop and sbottom searches usually assume a single decay mode dominates. As we saw in our earlier work~\cite{CahillRowley:2012rv}, specifically in the case of models with low fine-tuning, both 
stops and sbottoms have multiple decay modes with comparable branching fractions making their observation in any one channel significantly more difficult. This is further compounded 
by the multiple cascades that can take place due to the subsequent decays of the light electroweak gauginos (down to the LSP) into which the stops and sbottoms themselves decay. 

Figure~\ref{fig:EWGhist} shows the mass distributions for both the LSP and the lightest chargino which are seen to track each other quite closely overall. This is to be expected as the bulk of our 
pMSSM models have LSPs which are dominantly wino- or Higgsino-like and thus have a nearby charged partner. The 7 and 8 TeV searches (indirectly) cause almost all of the reduction 
in the model set and none of the other searches are seen to be particularly effective (except for HSCP) outside of the lowest mass bins. 
The effectiveness of the HSCP search is due to the existence of many HSCPs which are nearly LSP-degenerate charginos.



\begin{figure}
\centering
\subfloat{
     \begin{overpic}[height=3.5in]{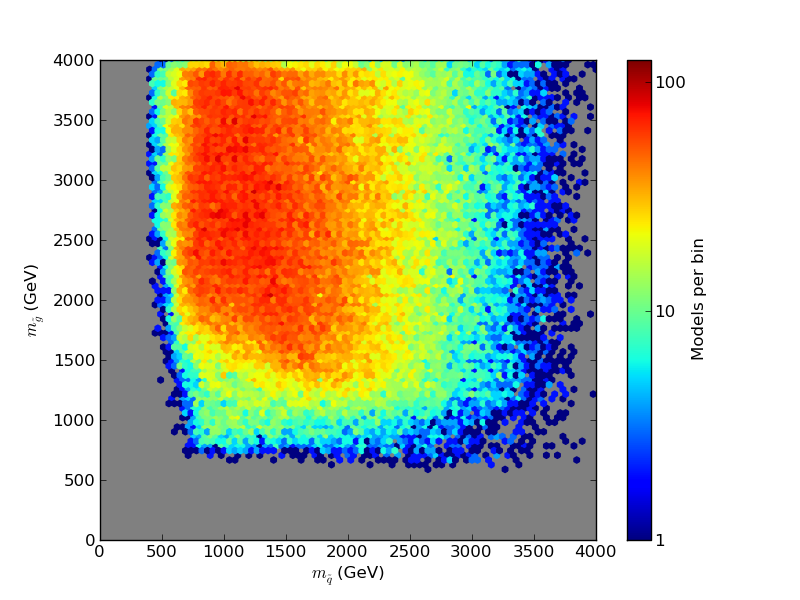}
     \end{overpic}
     } \\
\subfloat{
     \begin{overpic}[height=3.5in]{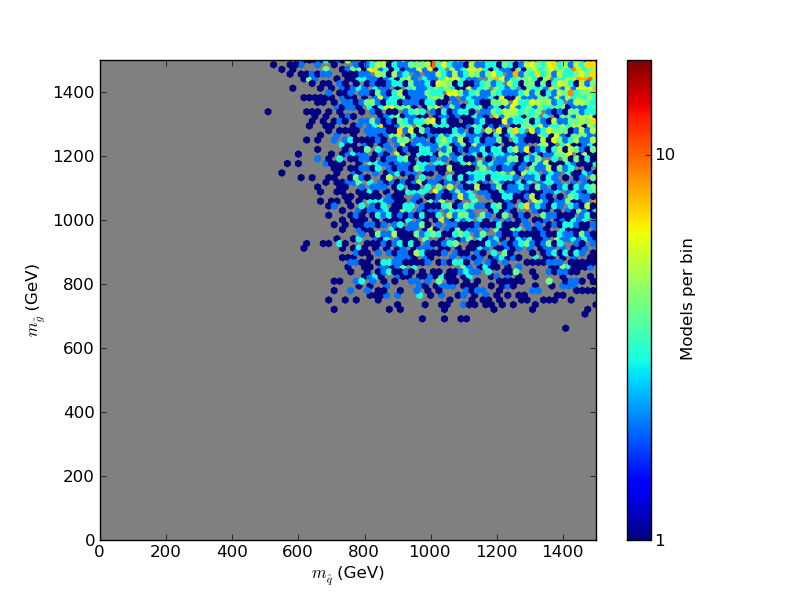}
     \end{overpic}
     }
\vspace*{0.5cm}
\caption{The density of models in the lightest 1st/2nd generation squark - gluino mass plane, after applying the LHC SUSY search constraints. The lower panel is a close-up of the low mass region of the upper panel.}
\label{fig:SGdensity}
\end{figure}

\begin{figure}
\centering
\subfloat{
     \begin{overpic}[width=3.5in]{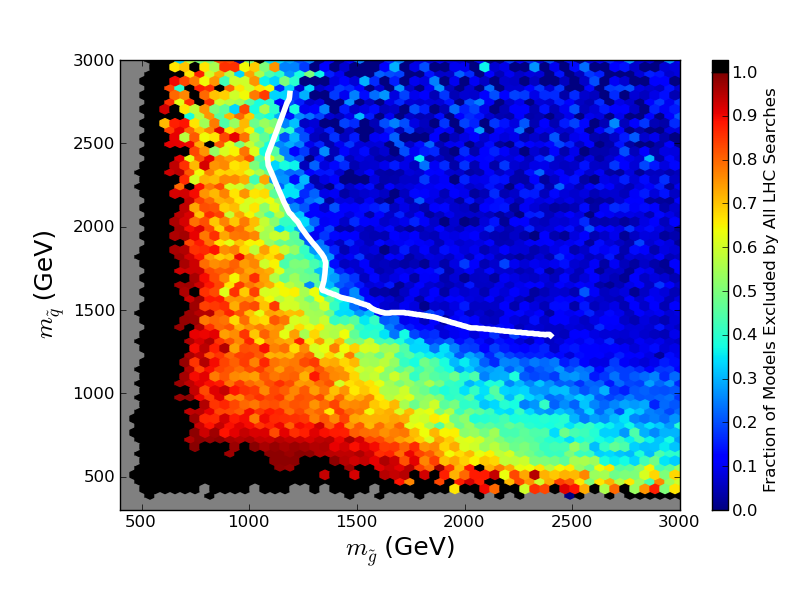}
     \end{overpic}
     }
\vspace*{0.5cm}
\caption{The efficiency of the LHC SUSY searches in excluding models in the lightest 1st/2nd generation squark - gluino mass plane. Bins for which all models have been excluded are colored black, while the simplified model limit from~\cite{ATLAS:2012ee} assuming degenerate squarks is shown in white.}
\label{fig:SGefficiency}
\end{figure}

Let us return to the models with relatively light squarks and gluinos which remain valid after all of the search constraints are applied. Figure~\ref{fig:SGdensity} displays the density of the gluino and lightest first/second generation squark masses after the full set of searches (except for the application of the Higgs 
mass cut which we now know does not have much of an effect except for lowering the overall statistics). The upper panel shows results for the entire model set distribution whereas 
the lower panel highlights only the squark and gluino mass range below 1.5 TeV. Recall that this is essentially the excluded region in the case of the cMSSM due to the ATLAS jets + MET generic searches. From this figure it is clear that the searches have predominantly excluded models with either light gluinos and/or light squarks with the gluinos showing greater sensitivity 
as expected. However, the zoomed-in lower panel demonstrates that models with relatively light squarks and gluinos do remain in the surviving model sample. In particular, 
models with gluinos below $\sim$700 GeV or squarks below $\sim$600 GeV remain viable. In addition, Figure~\ref{fig:SGefficiency} shows the fraction of models excluded by all LHC searches in the gluino-squark mass plane. As a guide, the simplified model limit from~\cite{ATLAS:2012ee} for eight degenerate squarks and a gluino is overlaid, and we see that colored sparticles lighter than this limit are quite viable. This is particularly true in the case of squarks, which need not be degenerate in the pMSSM.


\begin{figure}
\centering
\subfloat{
     \begin{overpic}[height=3.5in]{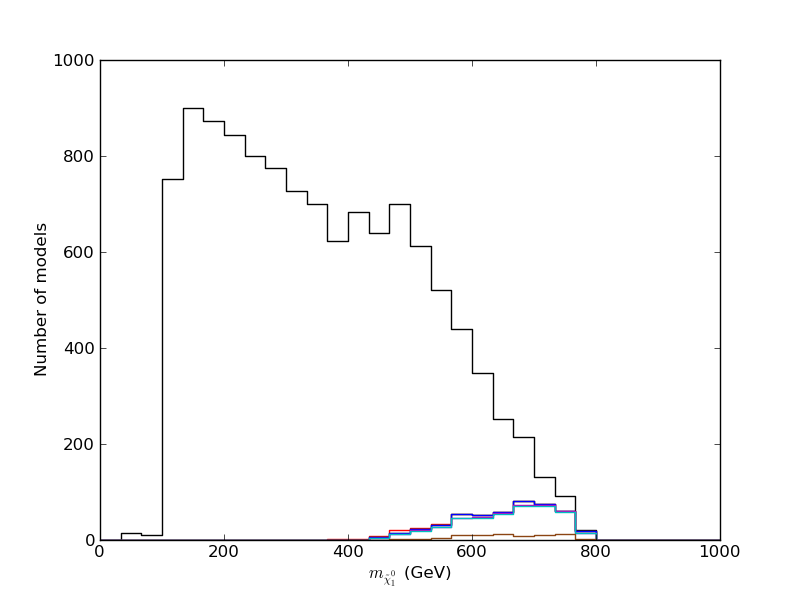}
     \end{overpic}
     } \\
\subfloat{
     \begin{overpic}[height=3.5in]{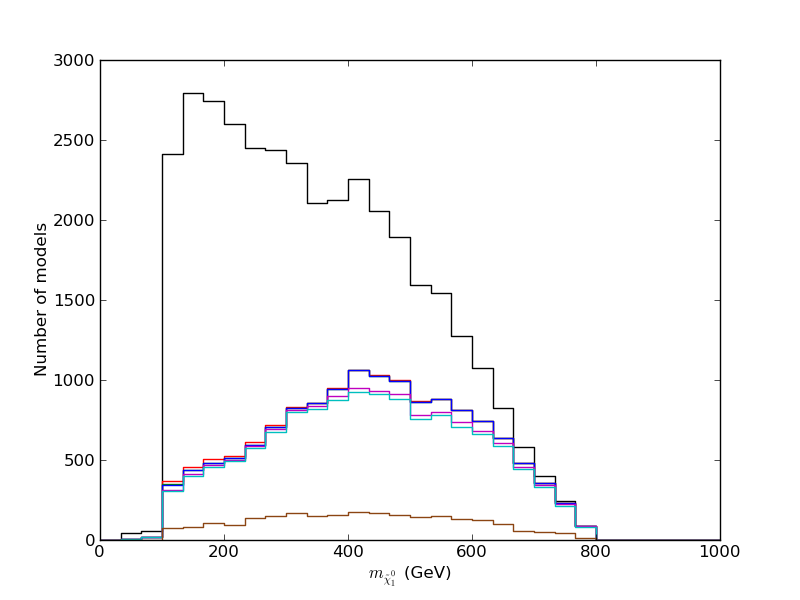}
     \end{overpic}
     }
\vspace*{0.5cm}
\caption{Histograms of the lightest neutralino mass in models with a gluino (upper panel) or lightest 1st/2nd generation squark (lower panel) with a mass below 800 GeV. The line colors are as in Figure~\ref{fig:SGhist}.}
\label{fig:Compressed}
\end{figure}

\begin{figure}
\centering
\subfloat{
     \begin{overpic}[height=3.5in]{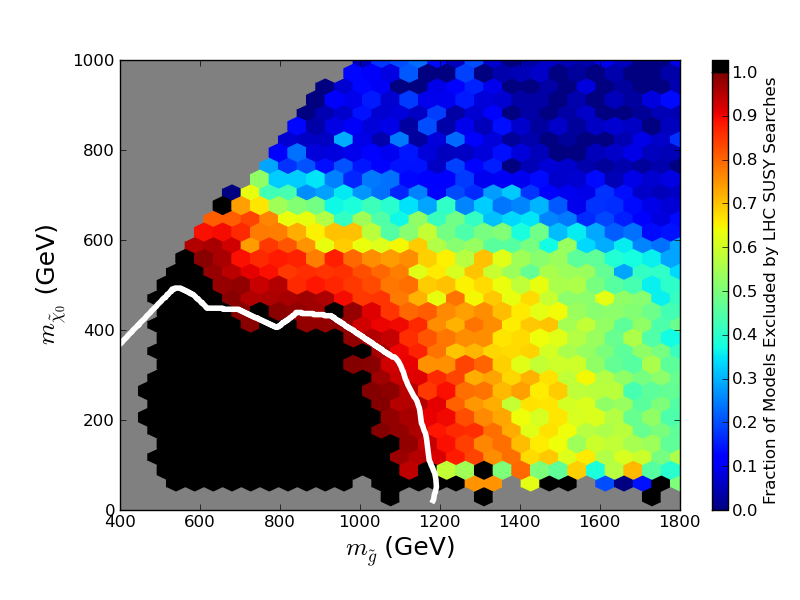}
     \end{overpic}
     }
\vspace*{0.5cm}
\caption{Fraction of models excluded in the gluino-LSP mass plane. The overlaid simplified model limit, denoted by the white curve, is taken from~\cite{ATLAS:2012ee}.}
\label{fig:LSPGluinoefficiency}
\end{figure}

\begin{figure}
\centering
\subfloat{
     \begin{overpic}[width=3.5in]{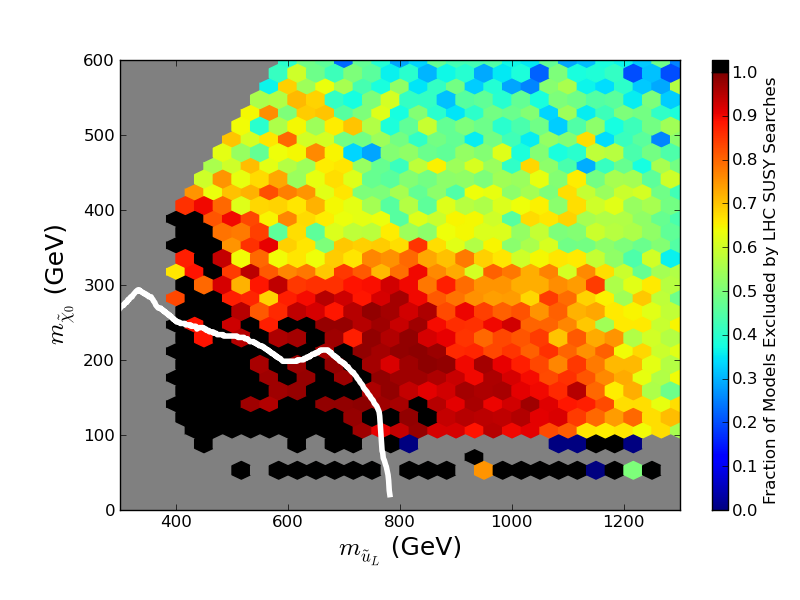}
     \end{overpic}
     } ~
\subfloat{ 
     \begin{overpic}[width=3.5in]{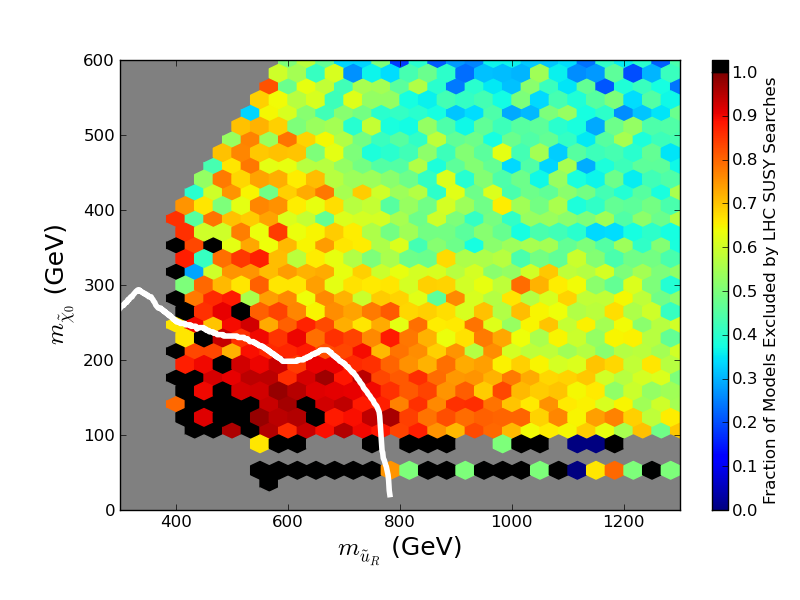}
     \end{overpic}
     } \\
\subfloat{
     \begin{overpic}[width=3.5in]{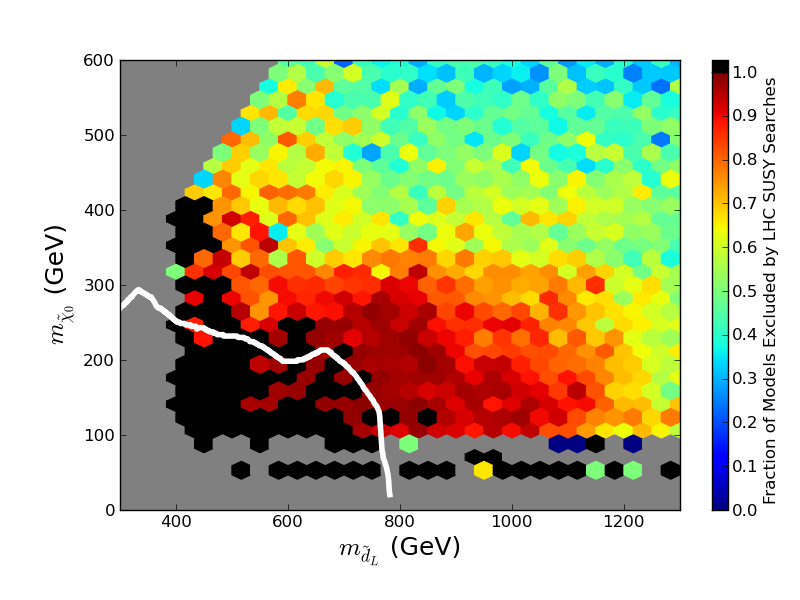}
     \end{overpic}
     } ~
\subfloat{
     \begin{overpic}[width=3.5in]{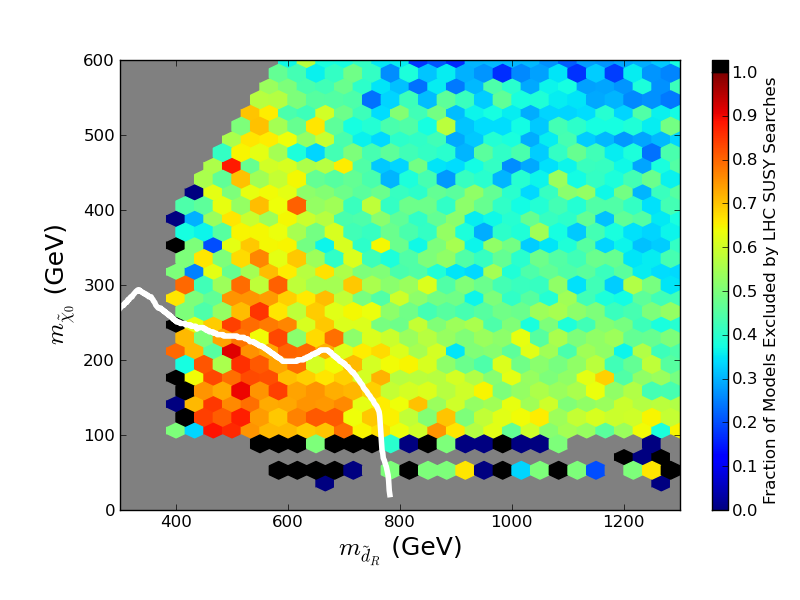}
     \end{overpic}
     }
\vspace*{0.5cm}
\caption{Fraction of models excluded in the $\tilde{u}_L$-LSP (upper left), $\tilde{u}_R$-LSP (upper right), $\tilde{d}_L$-LSP (lower left) and $\tilde{d}_R$-LSP (lower right) mass planes. The overlaid simplified model limit for eight degenerate squarks decaying to a neutralino LSP is taken from~\cite{ATLAS:2012ee}.}
\label{fig:LSPSquarkefficiency}
\end{figure}

To understand how these models survive the onslaught of the multiple LHC searches, we show in Figure~\ref{fig:Compressed} the mass distribution histograms for the LSP in the subset of survivor models for 
gluinos (top panel) or first/second generation squarks (bottom panel) with masses below 800 GeV as the various searches are imposed. One sees in both cases, after all the 
searches are combined (but primarily arising from just the 7 and 8 TeV general MET searches), that the LSP masses are quite peaked near the upper limit of their possible mass range. 
This demonstrates that the models with light squarks and/or gluinos that are survivors are mostly the result of compressed spectra, particularly in the case of light surviving gluinos. Furthermore, by looking at the exclusion efficiency in Figures~\ref{fig:LSPGluinoefficiency} and~\ref{fig:LSPSquarkefficiency}, we again see that models with compressed spectra are significantly less constrained than those with light LSPs. The left panels of Figure~\ref{fig:LSPSquarkefficiency} are very similar because the single parameter $m_{Q_1}$ sets the scale of both $\tilde{u}_L$ and $\tilde{d}_L$, up to electroweak corrections. However, since $\tilde{u}_R$ and $\tilde{d}_R$ are not necessarily degenerate in the pMSSM, the LHC's reach in the right panels of Figure~\ref{fig:LSPSquarkefficiency} is lower; the reach for right-handed down squarks is particularly poor as a result of PDF suppression. As a result, models with light right-handed squarks, particularly light $\tilde{d}_R$, may survive even if they have uncompressed spectra.

In some cases, models with light squarks or gluinos can be even more challenging to observe than their cross-section and spectrum compression would suggest, as a result of especially challenging decay patterns. For example, 86 models remain below the simplified model limit for gluinos decaying to $q \bar{q} \chi_0$ as shown in Figure~\ref{fig:LSPGluinoefficiency}. In all but 3 of these models, the dominant gluino decay mode results in 3rd generation squarks and/or quarks. The most common scenario is a gluino decaying to a Higgsino multiplet, in which case the decays are split between $\tilde{g} \to b \bar{b} \chi^0_{1,2}$, $\tilde{g} \to t \bar{t} \chi^0_{1,2}$, and $\tilde{g} \to b \bar{t} \chi^+_{1}$. The multiple final states present a challenge to any search targeting these models, which require optimization for a mixed final state including both tops and bottoms, rather than, e.g., 4 top quarks. In the remaining three models, the 3rd generation decays are subdominant but still significant, splitting the signal between the standard jets + MET searches and the gluino-mediated stop search. The situation is somewhat different when considering the surviving models with very light squarks in Figure~\ref{fig:LSPSquarkefficiency}. In this case, many models are expected to survive well below the simplified model limit, simply because fewer squarks are present than the simplified model assumes. However, it is still interesting to examine the characteristics of the models that are far below the simplified model limit. In these cases, the LSP is generally part of a Higgsino multiplet, with a heavier bino still being below the squark mass. Since the squark-Higgsino coupling is mass-suppressed, the decay to the heavier bino is dominant. The bino then decays to a charged or neutral Higgsino, resulting in on-shell or off-shell gauge bosons. While we would expect this decay to be seen in, e.g., the 8 TeV jets + lepton + MET search~\cite{ATLAS:2012gg}, if we look more closely we find that this search requires four jets with pT $>$ 80 GeV in addition to the hard lepton. Two of these jets may typically come from the squark $\to$ bino decay. However, the jets from the hadronic bino $\to$ Higgsino decays are frequently too soft to pass the 80 GeV cut, which both of them (or one and a jet from initial state radiation) must do since the other bino decays leptonically. Although models with light non-degenerate squarks are not as well motivated as those with light stops and sbottoms, they are interesting because they can have relatively large SUSY production cross-sections while remaining undetected by the present searches.

\begin{figure}
\centering
\subfloat{
     \begin{overpic}[height=3.5in]{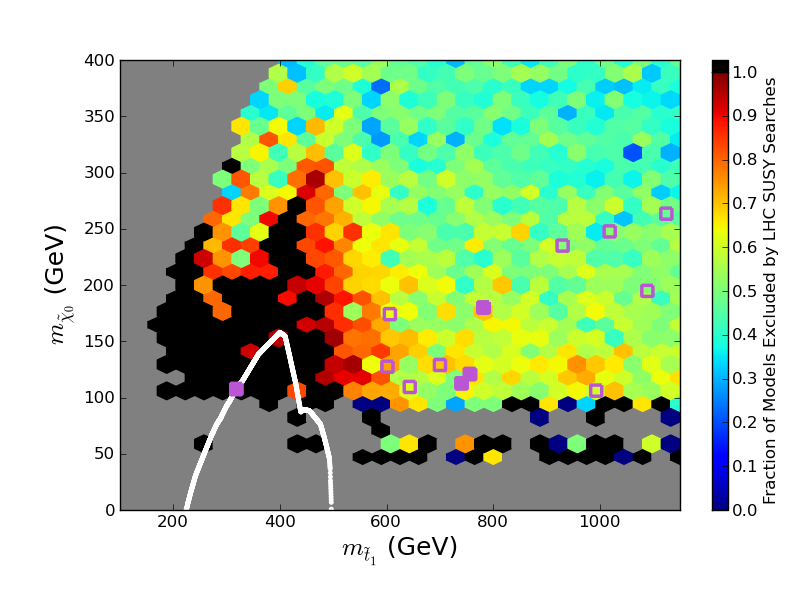}
     \end{overpic}
     } \\
\subfloat{
     \begin{overpic}[height=3.5in]{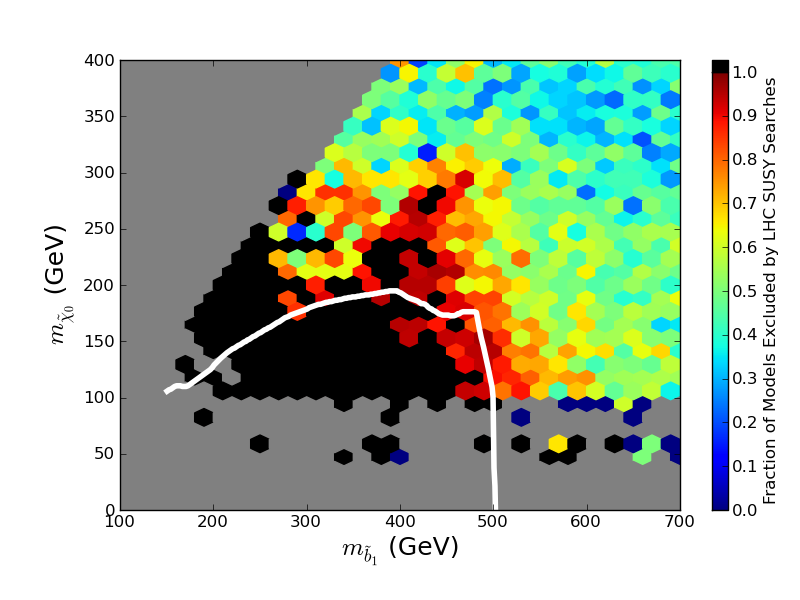}
     \end{overpic}
     }
\vspace*{0.5cm}
\caption{Fraction of models excluded in the $\tilde{t}_1$-LSP (upper) and $\tilde{b}_1$-LSP (lower) mass planes. The overlaid simplified model limits are taken from~\cite{stoplimit} and~\cite{ATLAS:directsbottom}. In the upper plane, the 13 models with $m_h = 126 \pm 3$ GeV and fine-tuning less than 1\% discussed in~\cite{CahillRowley:2012rv} are indicated with squares, which are filled (empty) for models that are excluded by (survive) all searches to date.}
\label{fig:3GLSPefficiency}
\end{figure}

We now turn our attention to models with light third-generation squarks, motivated by our previous discussion of fine-tuning~\cite{CahillRowley:2012rv}. It is interesting to ask if the general 8 TeV searches or any of the dedicated HF or ML searches are able to exclude some of the 13 low fine-tuning models{\footnote {Here we have 
employed the standard Barbieri-Giudice fine-tuning definition~\cite{Ellis:1986yg, Barbieri:1987fn}.}}, \ie, those with fine-tuning better than 1\%, with neutralino LSPs and $m_h=126\pm 3$ GeV as were discussed 
in our earlier work~\cite{CahillRowley:2012rv}. We find, in fact, that only one of these models (the one with the lightest stop and sbottom as one might expect) is actually excluded by the HF analyses, 
in particular, by both of the heavy stop searches (0l and 1l) and the direct sbottom search. Furthermore, we find that this same model, as well as an additional 3 of the remaining 12 models, are excluded 
by the ``vanilla'' MET searches at 8 TeV as light stops and sbottoms produce jets plus MET and these models have (in 2 out of 3 cases) relatively 
light gluinos. Figure~\ref{fig:3GLSPefficiency} shows the exclusion efficiency of the LHC SUSY searches for models with light third generation squarks, with the 13 low fine-tuning models highlighted in the upper panel. We see that only one of our low fine-tuning models lies in a red or black bin, where the current searches are highly effective (this is in fact the single model excluded by the HF searches), and that a dramatic increase in the reach of the HF searches will be necessary to fully explore this low fine-tuning region. If we consider the somewhat larger set of 50 neutralino LSP models satisfying a more relaxed fine-tuning constraint (the fine-tuning parameter $\Delta < 120$) with $m_h=126\pm 3$ GeV, we find that 
34 of them survive the 7 TeV general MET searches. 24 of these also survive the corresponding 8 TeV ``vanilla'' MET searches. If we now apply {\it all} of our LHC search constraints 
above we find that 22 models still survive at present. Two of the models are excluded by the non-MET searches. It is interesting that the 3$^{rd}$ generation searches are not having a large effect on this class of pMSSM models. This is partially the result of the stops and sbottoms simply being too heavy for the current searches. However, even low fine-tuning models with light stops or sbottoms can be challenging to exclude because of the array of possible decay channels available to the stop or sbottom, since most of these models have both wino and Higgsino multiplets below the stop mass.

\begin{figure}
\centerline{\includegraphics[height=6in]{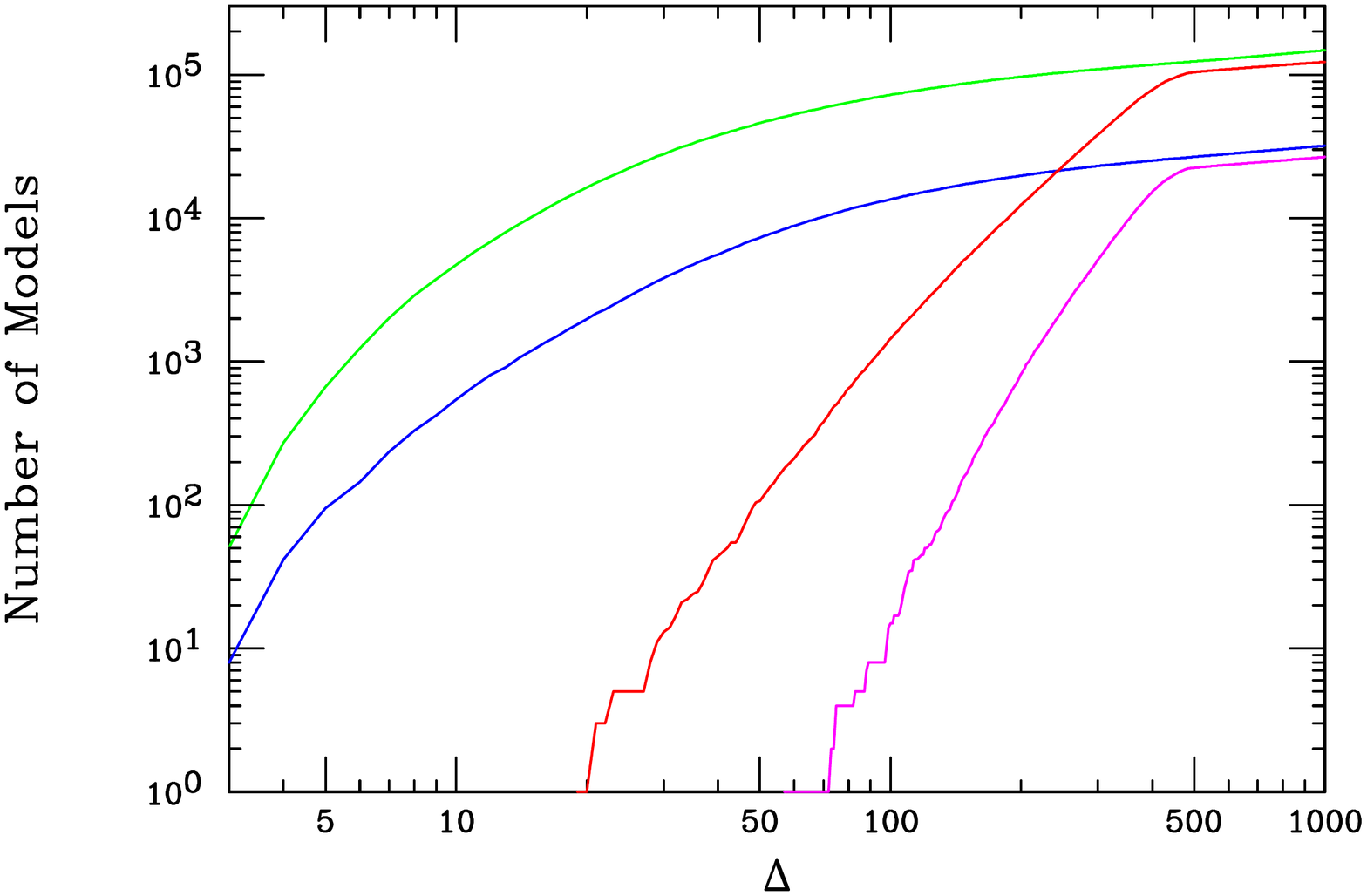}}
\vspace*{0.5cm}
\caption{Number of models with FT $\leq \Delta$ according to the fine-tuning measures of Barbieri and Giudice~\cite{Ellis:1986yg, Barbieri:1987fn} and Baer et. al~\cite{Baer:2012up,Baer:2012mv}. From top to bottom on the right side of the figure, the histograms are for the full neutralino LSP model set with the Baer et al. measure (green) and B-G measure (red), and the neutralino Higgs subset with the Baer et al. measure (blue) and B-G measure (magenta).}
\label{fig:ft3}
\end{figure}

Of course there are alternative ways of looking at fine-tuning. A recent example of this is provided by Refs.\cite{Baer:2012up,Baer:2012mv} which considered a fine-tuning measure that depends only 
upon weak scale quantities which is arguably what we are dealing with when discussing the pMSSM. The fine-tuning requirements in such a case are much looser than in the more traditional 
Barbieri-Giudice (B-G) measure so that many more models will have low fine-tuning employing this definition. Of course, the largest contributor to fine-tuning using either measure arises due to 
the size of the $\mu$, \ie, FT$_\mu \sim \mu^2/M_Z^2$. In Figure~\ref{fig:ft3} we compare the B-G measure with the electroweak fine-tuning measure presented by Baer et al. in Refs.\cite{Baer:2012up,Baer:2012mv} for our neutralino
pMSSM model set and for the Higgs subset. As can be seen in this figure, the number of low fine-tuning ($<$100) models using this weak scale measure is several orders of magnitude larger, even after the Higgs mass constraint 
is applied, with the Baer et al. measure. Furthermore, there are now a statistically reasonable subset of models with fine-tuning values $<$10. 

\begin{figure}
\centering
\subfloat{
     \begin{overpic}[height=3.5in]{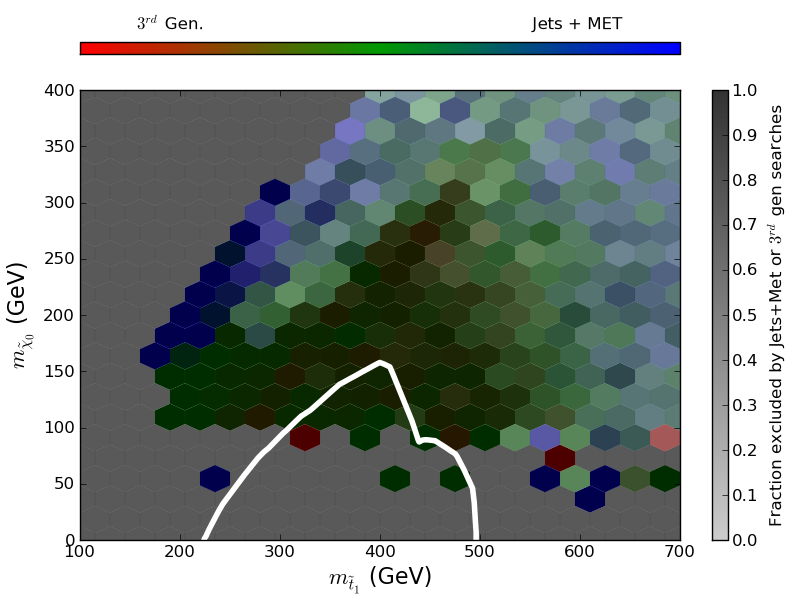}
     \end{overpic}
     } \\
\subfloat{
     \begin{overpic}[height=3.5in]{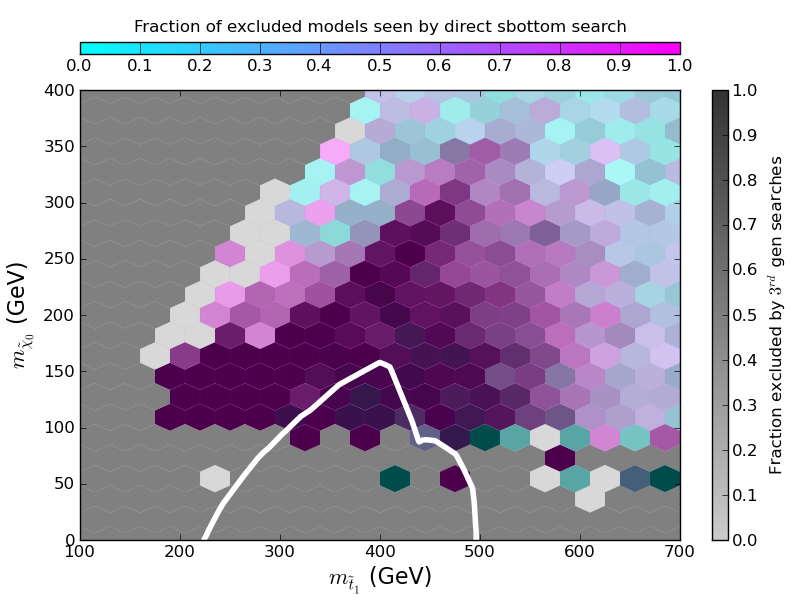}
     \end{overpic}
     }
\vspace*{0.5cm}
\caption{The efficiency of different LHC SUSY searches in the $\tilde{t}_1$-LSP mass plane. The overlaid simplified model limit is taken from~\cite{stoplimit}. Each bin in the upper panel is shaded with saturation according to the fraction of models excluded by the combination of jets + MET and third generation searches, and colored according to the search which is excluding the largest fraction of models in a given bin. In the lower panel, the coloration shows the fraction of all excluded models that are excluded by the direct sbottom search. We see that the direct sbottom search is quite effective in some regions of parameter space.}
\label{fig:3GLSPefficiency2}
\end{figure}

Returning to Figure~\ref{fig:3GLSPefficiency}, we see that the excluded regions of parameter space do not closely follow the simplified model limits. It is thus interesting to ask which searches are responsible for the majority of the search reach, and what decay patterns are easy or difficult to see with these searches. First, we can compare the performance of the standard jets + MET searches with the targeted HF searches, as shown in the upper panel of Figure~\ref{fig:3GLSPefficiency2}. We see that the third generation searches are ineffective for models that have a relatively small splitting between the stop and LSP ($\delta m \lesssim$ 130 GeV), so the exclusion of these models relies almost entirely on the standard jets + MET searches. In the uncompressed region, on the other hand, the third generation searches perform better, as expected. In particular, we note the presence of some red bins (where more models are seen by HF searches than by the ``vanilla'' seaches) with relatively uncompressed stops as light as 300 GeV, which demonstrate that the third-generation searches cover important gaps in the jets + MET coverage. We can also examine the reaches of the individual third-generation searches. As demonstrated in the lower panel of Figure~\ref{fig:3GLSPefficiency2}, the direct sbottom search~\cite{ATLAS:directsbottom} is responsible for nearly all of the HF exclusion power for models with relatively light stops (excluded models with heavy stops are generally seen by the gluino-mediated stop search~\cite{:2012pq}). We trace the surprising effectiveness of the direct sbottom search for models with light stops to the prevalence of wino and Higgsino LSPs in our model set. When the LSP is a wino or Higgsino, it is nearly degenerate with a chargino, and thus for light stops, the decay $\tilde{t}_1 \to b \tilde{\chi}_1^+$ has significantly more phase space than $\tilde{t}_1 \to t \tilde{\chi}_1^0$, and is often favored. The first of these decays typically leads to $b$-jets and missing energy, as the decay products of the chargino tend to be too soft to reconstruct. Since relatively hard $b$-jets are an easier signature to detect than top quarks (compare the simplified model limits in the upper and lower panels of Figure~\ref{fig:3GLSPefficiency}), the direct sbottom search has a much larger reach than the direct stop search, especially when the $\tilde{t}_1 \to b \tilde{\chi}_1^+$ mode is favored by kinematics. From the lower panel of Figure~\ref{fig:3GLSPefficiency2}, we see that the direct sbottom search loses efficiency for $\delta m \lesssim$ 130 GeV, which is surprising given that the $b$-jet reconstruction should be efficient down to at least 40 GeV (see for example Figure~\ref{fig:btag}). Looking at the cuts used in the direct sbottom search, we note that there is a region where the leading jet is not required to be a $b$-jet (relying instead on initial state radiation). However, this region still requires MET $>$ 150 GeV, which along with the requirement of a hard ISR jet (pT $>$ 130 GeV) is presumably responsible for the observed decrease in the effectiveness of this search in this region. 

Interestingly, only one model survives in the region below the simplified model limit for light stops; as indicated in Figure~\ref{fig:3GLSPefficiency}, it is at the very edge of the excluded region containing a 423 GeV stop and a 108 GeV LSP. From Figure~\ref{fig:3GLSPefficiency} it appears that the simplified model limit for stops is somewhat conservative; this is the result of decays to top quarks being a more difficult signature than direct decays to bottom quarks. Indeed, if we look at the models which survive with the lightest stops and largest stop-LSP splittings, we see that the stops either decay via $\tilde{t}_1 \to t \chi^0$ or produce a similar signature (a gauge boson and a relatively soft $b$-jet) by decaying through a heavier chargino $\tilde{t}_1 \to b \chi_2^+$, where the second chargino decay produces a W, Z, or Higgs (the single model inside the simplified model limit exhibits this decay pattern). The situation is reversed when we consider models with light sbottoms; now the simplified model limit is for the easiest scenario ($\tilde{b}_1 \to b \chi^0$), while the presence of charginos allows the decay $\tilde{b}_1 \to t \chi^+$. It is unsurprising, therefore, that in this case we find many allowed models below the sbottom simplified model limit. One particularly interesting case is a 406 GeV sbottom with a 49 GeV bino LSP. In this model, instead of decaying directly to the bino, the sbottom decays dominantly to a Higgsino multiplet at $\sim$185 GeV, with a 27\% branching fraction to $t \chi_1^+$. The Higgsinos decay to the bino and a Z, Higgs, or W boson in roughly equal proportions. Since the weak boson decays generally produce several jets, one might expect this model to be observed by the heavy stop search with 0 leptons~\cite{:2012si}. However, the sbottom also has a 17\% branching fraction to $b \chi_1^0$, in which case it produces too few jets to be detected by the stop search. This model provides a specific example of the general case we previously discussed in~\cite{CahillRowley:2012rv}, where the different decay modes of a sparticle are best seen by different searches, splitting the signal between multiple channels.

As we have seen, the third-generation searches are most sensitive to models in which stop or sbottom decays produce hard $b$-jets, and have a more difficult time observing models with stops or sbottoms that produce tops or decay to intermediate gauginos (resulting in softer $b$-jets). One potential way to increase sensitivity to these models could be a new direct stop search optimized for the mixed decay $\tilde{t}_1 \to t \chi^0$ + $\tilde{t}_1 \to b \chi^+$, a signature that is favored by combinatorics and can take advantage of decays that produce hard $b$-jets even when these decays are sub-dominant.

\begin{figure}
\centering
\subfloat{
     \begin{overpic}[height=3.5in]{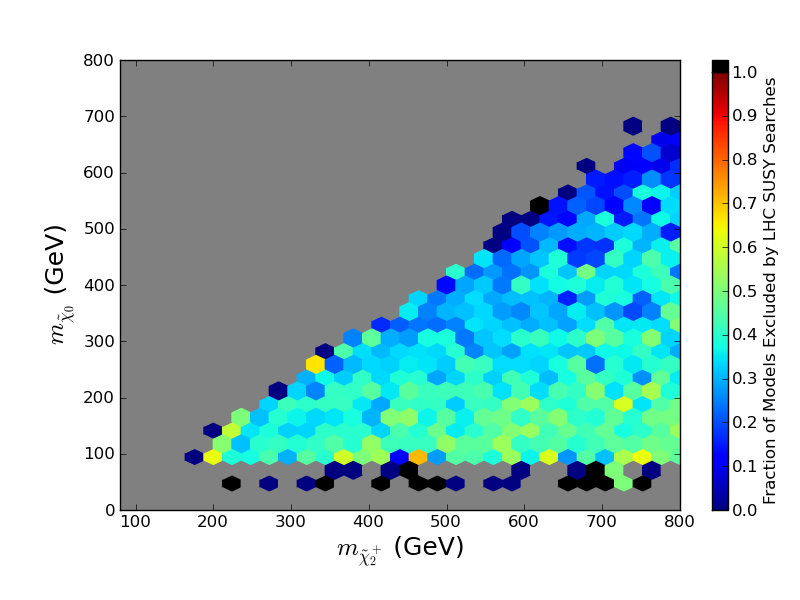}
     \end{overpic}
     }
\vspace*{0.5cm}
\caption{Fraction of models excluded in the $\tilde{\chi}_2^+$-LSP mass plane. We use $\tilde{\chi}_2^+$ rather than $\tilde{\chi}_1^+$ on the horizontal axis as $\tilde{\chi}_1^+$ and the LSP are nearly always degenerate.}
\label{fig:LSPCharginoefficiency}
\end{figure}

Finally, we can investigate the sensitivity of the current searches to the direct production of light gauginos. Figure~\ref{fig:LSPCharginoefficiency} shows the overall efficiency of the searches we have considered in the $\tilde{\chi}_2^+$-LSP mass plane. The relatively even coloration throughout the figure demonstrates that the multilepton searches designed to look for light gauginos (and sleptons) are not directly sensitive to models with light charginos, even with relatively uncompressed spectra. The main reason is that these searches are currently only powerful for nearly massless LSPs, whereas most of our LSPs are above 100 GeV due to the charged sparticle exclusion limit from LEP. Indeed, typical benchmarks for these searches consist of gauginos decaying to a single massless LSP. In our model set, not only is the LSP massive, but it also tends to be nearly degenerate in mass with some other gauginos in the case that the LSP is a wino/Higgsino. We encourage the continued consideration of benchmark scenarios with full wino and/or Higgsino multiplets at the bottom of the SUSY spectrum, as such possibilities are favored in our pMSSM scan by virtue of their efficient annihilation yielding a relic density below the WMAP bound.

It is hopeful that a high energy $e^+e^-$ collider will be built in the future to study the Higgs boson and other possible new physics discovered at the LHC. It is thus 
important to ask what SUSY particles, if any, remain kinematically accessible to such a collider with a given value of $\sqrt s$ accounting for the LHC search results. 
Table~\ref{LCimpact} presents the fraction of our pMSSM models that survive all LHC constraints (including that for the Higgs mass) which have a given sparticle kinematically accessible 
via pair production for different possible values of $\sqrt s$ for such a high energy $e^+e^-$ collider, while Table~\ref{LCimpact2} shows the efficiencies of the LHC searches on the models with kinematically accessible sparticles. For $\sqrt s=250$ GeV, which is appropriate for a Higgs factory, there is a small sample of models for which the 
first two neutralinos and the lightest chargino are accessible; this fraction grows substantially for $\sqrt s=500$ GeV. These results should not be too surprising 
since most of our LSPs are either wino- or Higgsino-like. This suggests that light gauginos will be the most likely target of a low energy $e^+e^-$ collider given the 
present LHC data. This result is, of course, further reinforced by low fine-tuning considerations. In the B-G approach, the largest tree-level fine-tuning contribution arises from $\mu$,  
FT$_\mu \simeq 4\mu^2/M_Z^2$. Thus requiring fine-tuning $<$ 100 (10) implies that $|\mu| < 456$ (144) GeV so that the light Higgsino-like states should be anticipated in the low-mass part of 
the spectrum.

\begin{table}
\centering
\begin{tabular}{|l|c|c|c|c|} \hline\hline
Sparticle     &  $\sqrt s=250$ GeV  & $\sqrt s=500$ GeV & $\sqrt s=1$ TeV & $\sqrt s=2$ TeV \\ \hline
$\tilde \chi^0_{_1}$   &   3.3\% &  22.6\% &  58.1\% &  94.7\% \\
$\tilde \chi^0_{_2}$   &   1.3\% &  11.1\% &  30.0\% &  57.5\% \\
$\tilde \chi^0_{_3}$   &       &   0.7\% &   7.2\% &  25.5\% \\
$\tilde \chi^\pm_{_1}$ &   2.9\% &  22.2\% &  57.5\% &  93.6\% \\
$\tilde \chi^\pm_{_2}$ &       &   0.2\% &   4.8\% &  19.1\% \\
$\tilde e_L $        &         &   0.1\% &   2.2\% &  12.7\% \\
$\tilde e_R $        &         &   0.3\% &   3.2\% &  14.5\% \\
$\tilde \nu_e$         &         &   0.1\% &   2.3\% &  12.8\% \\
$\tilde \tau_1$      &         &   0.4\% &   4.7\% &  24.7\% \\
$\tilde \nu_\tau $   &         &   0.2\% &   2.4\% &  13.2\% \\
$\tilde t_1$         &         &         &   0.3\% &  10.4\% \\
$\tilde b_1$         &         &         &   0.8\% &  16.7\% \\
\hline\hline
\end{tabular}
\caption{Fraction (in percent) of viable pMSSM models with $m_h = 126 \pm 3$ GeV after the LHC searches that have kinematically accessible sparticles at an $e^+ e^-$ linear collider with various center of mass energies. The lack of an entry signals that the fraction is below 0.1\%.}
\label{LCimpact}
\end{table}

\begin{table}
\centering
\begin{tabular}{|l|c|c|c|c|} \hline\hline
Sparticle     &  $\sqrt s=250$ GeV  & $\sqrt s=500$ GeV & $\sqrt s=1$ TeV & $\sqrt s=2$ TeV \\ \hline
$\tilde \chi^0_{_1}$   &   48.3\% &  54.5\% &  60.5\%  &  67.3\%  \\
$\tilde \chi^0_{_2}$   &   53.0\% &  58.7\% &  64.6\% &  69.5\% \\
$\tilde \chi^0_{_3}$   &         &   61.9\% &  66.5\% &  69.0\% \\
$\tilde \chi^\pm_{_1}$ &   47.6\%  &  54.4\%  &  60.4\% &  67.3\% \\
$\tilde \chi^\pm_{_2}$ &         &   54.0\% &  66.5\% &  68.3\% \\
$\tilde e_L $        &         &   31.6\% &   56.5\% &  63.6\% \\
$\tilde e_R $        &         &  48.6\% &  58.2\% &  62.8\% \\
$\tilde \nu_e$         &         &  32.0\%  &  56.7\% &  63.7\% \\
$\tilde \tau_1$      &         &   48.0\%  &  58.6\% &  64.3\% \\
$\tilde \nu_\tau $   &         &  44.7\%  &   58.7\% &  64.5\% \\
$\tilde t_1$         &         &         &   27.8\% &  56.5\% \\
$\tilde b_1$         &         &         &   27.9\% &  58.3\% \\
\hline\hline
\end{tabular}
\caption{Ratio of the viable models after the LHC searches to the number of models in our sample before the LHC constraints are applied, scaled to the kinematic reach of the machine and requiring $m_h = 126 \pm 3$ GeV.  This represents the efficiency of the LHC searches, e.g., 54.4\% of the models in our sample that had $m_{\tilde\chi_1^\pm}<250$ GeV remain viable after the LHC search constraints are applied.}
\label{LCimpact2}
\end{table}

At $\sqrt s=500$ GeV, we begin to see small samples of models where both sleptons and the next heaviest set of electroweak gauginos become kinematically accessible, 
although the numbers are still quite small. Once $\sqrt s=1$ TeV is reached, small samples of models with kinematically accessible light stops and sbottoms are present. 
It is clear, however, that a sizable fraction of the sparticle spectrum will only become accessible at collision energies above $\sim$2 TeV.

\begin{figure}
\centering
\subfloat{
     \begin{overpic}[width=3.5in]{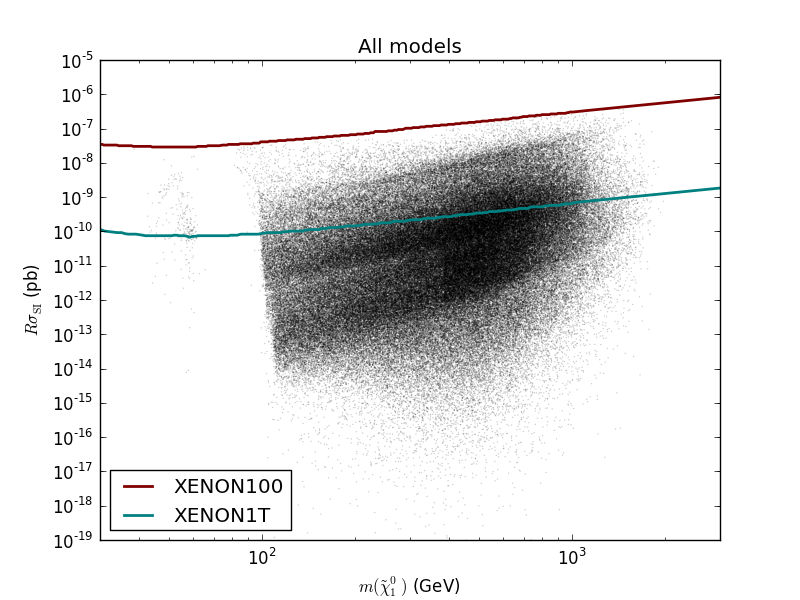}
     \end{overpic}
     } ~
\subfloat{ 
     \begin{overpic}[width=3.5in]{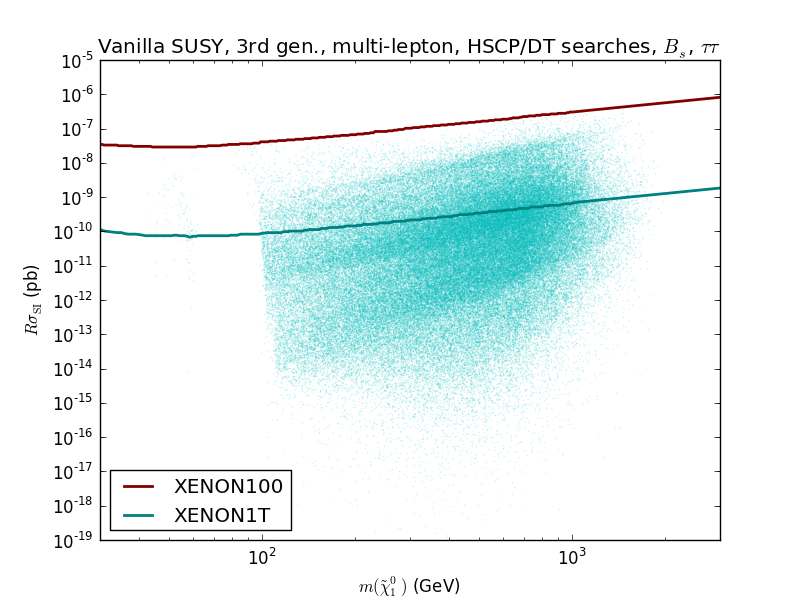}
     \end{overpic}
     } \\
\subfloat{
     \begin{overpic}[width=3.5in]{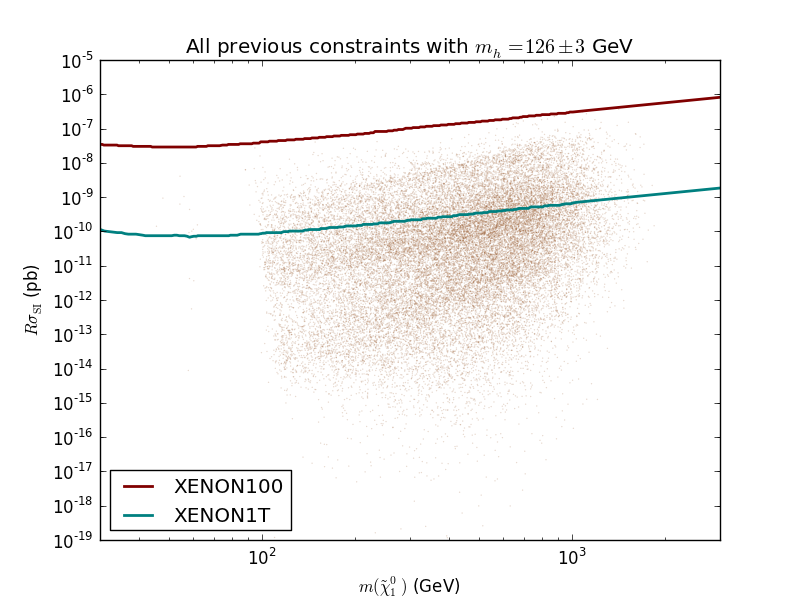}
     \end{overpic}
     } ~
\subfloat{
     \begin{overpic}[width=3.5in]{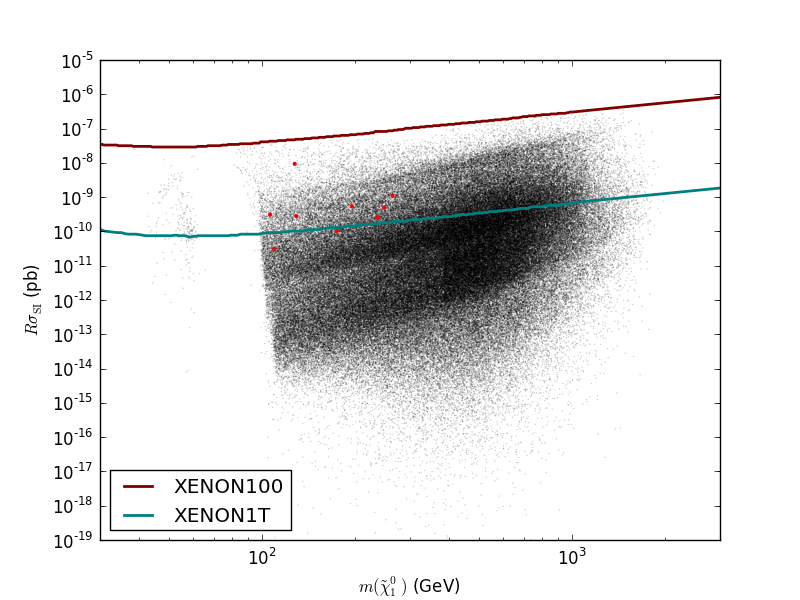}
     \end{overpic}
     }
\vspace*{0.5cm}
\caption{Scatter plots showing the mass of the lightest neutralino versus the spin-independent cross section for: all models (upper left), models which survive the LHC SUSY searches (upper right), models which survive the LHC SUSY searches and have m$_h$ = $126 \pm 3$ GeV (lower left), and models with low fine-tuning (lower right), which correspond to the red points.}
\label{fig:direct}
\end{figure}

Our results also have implications for dark matter searches. For example, Figure ~\ref{fig:direct} shows the direct detection spin-independent cross section (scaled by the ratio of the relic density to the total dark matter density) as a function of the LSP mass, and displays the recent XENON-100 bound~\cite{Aprile:2011hi} as well as the expected reach of XENON-1T~\cite{Xenon1T} for reference. The upper left (black) scatter plot shows the results for 
our original pMSSM neutralino model set as generated, while the upper right panel shows the corresponding subset of models surviving all the LHC search constraints except for 
the Higgs mass requirement. The lower left (brown) panel includes this constraint.  We see from this figure that, apart from the overall density of 
points, the LHC searches do not have much influence on the overall nature of the distribution of cross section predictions. The Higgs mass requirement, though thinning the point 
density significantly, is also seen to have little impact on these expectations. Thus we conclude that pMSSM searches at the LHC have very little impact on the possible range of 
spin-independent relic density-scaled cross sections, and that indeed these searches are complementary. The lower right panel shows the predicted relic density-scaled cross sections for the 
remaining low fine-tuning models discussed above. Here we see that these models should mostly be accessible to the XENON-1T experiment.


\section{Conclusion}
\label{sec:conc}

With the remarkable performance of the LHC, many searches for supersymmetry have been performed this year with both 7 TeV and 8 TeV data. It is important to ensure that as analyses are designed to search for new physics, experimental collaborations strive to obtain maximal coverage of the vast parameter space of SUSY. The most simple searches undoubtedly leave corners of this space unexplored, necessitating the identification of such regions and the development of new search strategies to survey them. Here, we have continued our previous work in this area by studying the capabilities of the latest LHC searches to constrain a large set of pMSSM models. Compared to previous searches, we find that significantly improved coverage of the pMSSM is afforded by the additional analyses. While our original model set did not require a Higgs mass near 126 GeV, our results are found to be generally independent of this additional constraint.

The first searches for SUSY with 8 TeV data employ signal regions that are very similar to those in their 7 TeV counterparts. With slightly tighter cuts and several downward fluctuations in the numbers of observed versus expected events, these searches successfully exclude models that had remained viable after the corresponding 7 TeV analyses. An increase in integrated luminosity at 8 TeV, however, will not cause the coverage of these searches to improve much further, unless either downward fluctuations of events continue to occur or reductions in the background uncertainties can be achieved. We also caution that a small fraction of models that are allowed by the 8 TeV searches had produced enough signal events to be excluded by the analogous 7 TeV searches. This result is symptomatic of the unavoidable trade-off between obtaining search sensitivity in new areas of parameter space and retaining sensitivity in previously probed regions.

In addition, directed searches for stops/sbottoms, sleptons, and electroweak gauginos involve less conventional signatures of supersymmetry, particularly $b$-jets and leptons in the final state. While these searches rule out fewer pMSSM models overall than the more general SUSY searches, they enjoy success in excluding some spectra that are missed by the latter. Often, such models have heavier gluinos and 1st/2nd generation squarks, making it difficult for conventional searches to see them. Because most of our gauginos are wino and Higgsino multiplets with very small mass splittings, the heavy flavor signal regions have comparatively better sensitivity to our pMSSM models than the direct gaugino searches.

The last class of searches that we have included in this work involves signatures without missing energy. Given the significant population of models with stable charginos within our model set, it is not surprising that searches for heavy stable charged particles and disappearing tracks exclude significant fractions of the models. Updates to the $B_s\to \mu^+\mu^-$ and $A,H\to \tau^+\tau^-$ analyses have improved the LHC coverage of our model set as well. Notably, since the signal regions for these non-MET searches involve criteria that are very different from those of other searches, the coverage gained from these analyses is nearly orthogonal to that obtained from the MET-based searches.

After including all of these analyses, we find that many pMSSM spectra with light sparticles survive. While supersymmetry continues to be limited by the lack of collider signatures thus far, the state of the full MSSM is far better than suggested by limits on simplified models or more constrained scenarios such as mSUGRA. In particular, we find surviving models containing 1st/2nd generation squarks, gluinos, and 3rd generation squarks with masses below 600, 700, and 400 GeV, respectively. Models with light colored sparticles that are not excluded by LHC searches generally have compressed spectra. We also observe viable models with fine-tuning near 1\%, indicating that the MSSM still has significant potential to provide a natural solution to the hierarchy problem.

Finally, we have considered the potential of experiments outside the LHC to observe supersymmetry. A linear collider would be able to produce many of the light gaugino states in our viable models. We emphasize that if the MSSM is indeed natural, light Higgsinos will be copiously produced at a future $e^+ e^-$ collider. Astrophysical experiments provide a complementary method of searching for supersymmetry by seeking signatures of dark matter, which is composed of neutralinos in the models considered here. By searching for direct interactions of the LSP with matter, an upgraded XENON experiment would probe a significant fraction of our pMSSM model set. Both of these avenues of exploration are orthogonal to the LHC supersymmetry searches, and would be particularly effective in probing natural SUSY.

Having surveyed the vast landscape of minimal supersymmetry with the pMSSM, we declare it to be alive and well. We look forward to upcoming collider and astrophysical results, in the hope of finally revealing the hitherto elusive face of SUSY.

\section*{Acknowledgments}

The authors are grateful to John Conway for providing an updated version of PGS, and for discussions with J.~Conley, R.~Cotta, T.~Eifert, J.~Gainer, M.~P.~Le, F.~Paige, and T.~Plehn. 

\newpage

%
\def\IJMP #1 #2 #3 {Int. J. Mod. Phys. A {\bf#1},\ #2 (#3)}
\def\MPL #1 #2 #3 {Mod. Phys. Lett. A {\bf#1},\ #2 (#3)}
\def\NPB #1 #2 #3 {Nucl. Phys. {\bf#1},\ #2 (#3)}
\def\PLBold #1 #2 #3 {Phys. Lett. {\bf#1},\ #2 (#3)}
\def\PLB #1 #2 #3 {Phys. Lett. B {\bf#1},\ #2 (#3)}
\def\PR #1 #2 #3 {Phys. Rep. {\bf#1},\ #2 (#3)}
\def\PRD #1 #2 #3 {Phys. Rev. D {\bf#1},\ #2 (#3)}
\def\PRL #1 #2 #3 {Phys. Rev. Lett. {\bf#1},\ #2 (#3)}
\def\PTT #1 #2 #3 {Prog. Theor. Phys. {\bf#1},\ #2 (#3)}
\def\RMP #1 #2 #3 {Rev. Mod. Phys. {\bf#1},\ #2 (#3)}
\def\ZPC #1 #2 #3 {Z. Phys. C {\bf#1},\ #2 (#3)}

\end{document}